\documentclass[preprint,12pt,authoryear]{elsarticle}
\usepackage[T1]{fontenc}
\usepackage[utf8]{inputenc}

\newcommand{\equatref}[1]{Eq.\ \eqref{#1}}                                              
\newcommand{\figref}[1]{Figure\ \ref{#1}}                                                
\usepackage{amssymb}
\usepackage{amsmath} 
\usepackage{amsfonts}                                                                   
\usepackage{bm}                                                                        
\usepackage{eucal}                                                                      

\usepackage{booktabs}  
\usepackage{subcaption} 

\usepackage{xcolor} 

\journal{Elsevier}

\begin{document}

\begin{frontmatter}

\title{Foliar Uptake of Biocides: Statistical Assessment of Compartmental and Diffusion-Based Models}

\author[inst1]{Enrico Sangoi}

\affiliation[inst1]{organization={Department of Chemical Engineering, University College London},
            addressline={Torrington Place}, 
            city={London},
            postcode={WC1E 7JE}, 
            state={United Kingdom},
            country={}}

\author[inst2]{Federica Cattani}
\author[inst3]{Faheem Padia}
\author[inst1]{Federico Galvanin}

\affiliation[inst2]{organization={Process Studies Group, Syngenta, Jealott's Hill International Research Centre},
            addressline={Berkshire}, 
            city={Bracknell},
            postcode={RG42 6EY}, 
            state={United Kingdom},
            country={}}
\affiliation[inst3]{organization={Formulation Technology Group, Syngenta, Jealott's Hill International Research Centre},
            addressline={Berkshire}, 
            city={Bracknell},
            postcode={RG42 6EY}, 
            state={United Kingdom},
            country={}}
            
\begin{abstract}
The global population increase leads to a high food demand, and to reach this target products such as pesticides are needed to protect the crops. Research is focusing on the development of new products that can be less harmful to the environment, and mathematical models are tools that can help to understand the mechanism of uptake of pesticides and then guide in the product development phase. This paper applies a systematic methodology to model the foliar uptake of pesticides, to take into account the uncertainties in the experimental data and in the model structure. A comparison between different models is conducted, focusing on the identifiability of model parameters through dynamic sensitivity profiles and correlation analysis. Lastly, data augmentation studies are conducted to exploit the model for the design of experiments and to provide a practical support to future experimental campaigns, paving the way for further application of model-based design of experiments techniques in the context of foliar uptake.
\end{abstract}


\end{frontmatter}


\section{Introduction}
\label{sec:intro}

As the world's population continues to grow and the planet's resources remain limited, ensuring sufficient food production becomes a crucial challenge both in the present and for the coming decades. In tackling this issue, the development of improved and safer biocides will be essential to optimize crop yields and meet the increasing demand for food. This brings forth the need for innovative solutions that align with sustainable agricultural practices (Umetsu and Shirai, \citeyear{umetsu_development_2020}). 

Crop protection products such as herbicides and pesticides can be delivered to the plants via different methods, with the spraying on the foliage being one of the most relevant ones for field applications. 
The process that bring the active ingredient (AI) from the mixing tank to the biological target sites is determined by a series of inter-correlated processes in the biodelivery chain. Having a quantitative understanding of these processes and their effect on the product efficacy is fundamental for developing innovative solutions, and mathematical models are tools that can help researchers in this field and guide further experimental directions.
While essential for crop protection and beneficial for tackling food demand needs, the use of pesticides raises concerns about environmental impact,  particularly in terms of soil and water contamination (Aktar et al., \citeyear{aktar_impact_2009}), as well as the impact on global ecosystems (Sharma et al., \citeyear{sharma_worldwide_2019}). 
This underscores the heightened importance of developing tools and technologies that can contribute to the production of safer pesticides for the surrounding environment.

Among all the processes identified in the biodelivery chain, the foliar uptake of the AI \citep{franke_mechanisms_1967}, i.e. the process of absorption of the AI through the leaves \citep{fernandez_foliar_2021}, is not completely understood and influenced by several factors, while being a crucial step in the path that leads the AI from the tank to the target sites, i.e. the biological macromolecules essential for the physiological functions of pests, weeds, and pathogens, that interact with the biocide \citep{zhang_how_2025}.
Therefore more effort both theoretical and experimental is needed to characterize this phenomenon subject to high uncertainty in its description.

Several works can be found in literature tackling the question of how to describe the foliar uptake process. The models available in the literature can be divided in three categories \citep{trapp_plant_2004}: \textit{i}) empirical correlations, \textit{ii}) compartmental models, and \textit{iii}) diffusion-based models. 
Examples of empirical correlations can be found in \cite{briggs_physico-chemical_1987} and \cite{forster_mechanisms_2004}, however these models are unable to describe the underlying mechanism. Compartmental models have been applied to physiological systems for several decades \citep{rowland_clearance_1973}, with specific applications also to foliar uptake, e.g. in the works by \cite{bridges_compartmental_1974}, \cite{satchivi_nonlinear_2000} and \cite{fantke_dynamics_2013}. 
Some works in literature \citep{schreiber_review_2006} suggest that the process of AI uptake through the cuticle, i.e. the outermost layer of leaves, can be by diffusion. Starting from this consideration, diffusion-based mechanistic models have been proposed in literature for the characterization of foliar uptake, e.g. in \cite{mercer_simple_2007} and \cite{tredenick_nonlinear_2017}.
To the best of our knowledge, there is no work proposed in literature where a systematic approach is applied for the development and statistical assessment of foliar uptake models.

The objective of this study is to obtain and statistically validate a predictive model to represent the phenomena occurring within the leaves in a quantitative way. 
To model biological systems, uncertainty typically arises in the experimental data and in the definition of a suitable structure of the model, i.e. in which phenomena should be included in the mathematical formulation.
Since the foliar uptake case study involves biological systems, the large uncertainty in the experimental observations must be taken into consideration when assessing the reliability of the mathematical models in a statistically sound approach.

This paper approaches the problem with a systematic modeling framework presented in Section~\ref{sec:methods}. Different models, described in Section~\ref{sec:models}, are considered for the characterization of foliar uptake. The results of their comparison are presented in Section~\ref{sec:results}, focusing on the identifiability of model parameters and data augmentation studies conducted to exploit information from the models for the design of additional experiments.
Section~\ref{sec:conclusions} summarizes the achievements of this study and points the direction of future works.

\section{Methodology}
\label{sec:methods}

\begin{figure}
    \centering
    \includegraphics[width=1.0\linewidth]{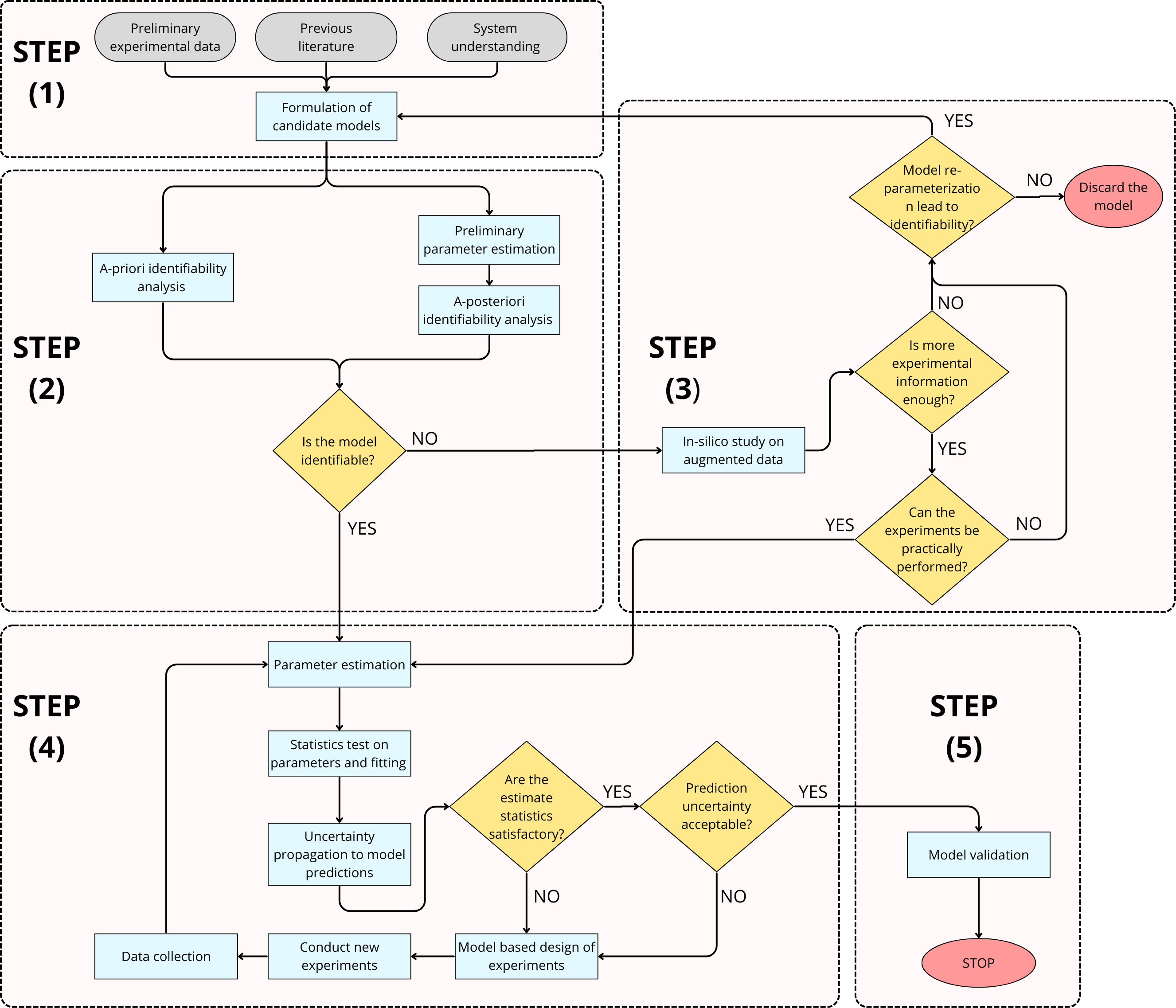}
        \caption{Framework considered to develop a predictive foliar uptake model.}
    \label{fig:framework}
\end{figure}

The general modeling framework considered in this study to obtain a reliable predictive model for the characterization of foliar uptake is presented in Fig.~\ref{fig:framework}.
The first step in the procedure is to formulate a set of candidate models, which can be based on previous literature available, the understanding of the physico-chemical processes involved in the system and preliminary experimental observations. 
The general formulation of a dynamic model involving differential and algebraic equations is the following
\begin{align} \label{eq:generalmodel}
\begin{cases}
&     \bm{\dot x}(t) = \bm{f}(\bm{x}(t),\bm{u}(t),\bm{\theta}, t) \\
&    \hat{\bm{y}}(t) = \bm{g}(\bm{x}(t),\bm{u}(t),\bm{\theta})
\end{cases}
\end{align}
where $t$ is the variable time, $\bm{x}(t)$ is a $N_x$-dimensional vector of system state variables, $\bm{\dot x}(t) $ the vector of time derivatives, $\bm{u}(t)$ the $N_u$-dimensional vector of known system inputs, $\bm{\theta}$ the $N_\theta$-dimensional vector of model parameters, and $\hat{\bm{y}}(t)$ the $N_y$-dimensional vector of predicted system outputs.

Once the candidate models are formulated, it is important to test the identifiability of their parameters (step 2 in Fig.~\ref{fig:framework}), i.e. if the model parameters can be uniquely identified from a given set of input and output measurements. Identifiability methods can be distinguished between \textit{a-priori} and \textit{a-posteriori} tests \citep{miao_identifiability_2011}: \textit{a-priori} methods consider uniquely the structure of the model, while \textit{a-posteriori} techniques start from preliminary experimental data and include practical experimental limitations.

The problem of model identifiability is expressed as
\begin{equation}\label{eq:id}
    \hat{\bm{y}}(\bm{\theta}_1) = \hat{\bm{y}}(\bm{\theta}_2) \Rightarrow \bm{\theta}_1=\bm{\theta}_2 ,
\end{equation}
meaning that if the model predictions $\hat{\bm{y}}$ are identical for some parameter vectors $\bm{\theta}_1$ and $\bm{\theta}_2$, then these vectors must be the same, i.e. $\bm{\theta}_1=\bm{\theta}_2$. 
If the condition in \equatref{eq:id} does not hold, then the model is not-uniquely identifiable, i.e. there exist two distinct vectors $\bm{\theta}_1$ and $\bm{\theta}_2$ which give the same model predictions.

If the model is not identifiable a-posteriori, before discarding it, other questions are posed in the proposed scheme (step 3). The first question is whether having additional experimental observations would be sufficient to solve the identifiability issues. To answer this question, an in silico study is conducted by simulating new experimental data and subsequently evaluating the expected improvement in the statistical quality of the estimates to justify the need for additional data. Moreover, it must be considered that in a real case application experiments will also have to be conducted in practice, which is another important decision block. If the outcome of these decision is still negative, model re-parametrization methods can be considered to solve the identifiability issues before removing a model from the set of candidate ones.

The modeling procedure then continues (step 4) with parameter estimation and statistical tests to assess both the quality of fitting and the precision on parameter estimation. Given that the final objective is to apply the model in practice, the next step is to propagate the uncertainty from parameters to model predictions, to assess whether this uncertainty is within an acceptable range. If the outcome of statistical tests and uncertainty propagation is not satisfactory, model-based design of experiments (MBDoE) can be applied to optimally design additional experiments, and the additional experimental evidence will be used to re-estimate the model parameters \citep{franceschini_model-based_2008}. Finally, the modeling procedure is concluded (step 5) by validating the model on independent experimental observations.

The following subsections will present in detail the methodology employed for testing model identifiability and for the in silico data augmentation study, which are the focus of the results presented in this paper.

\subsection{Identifiability analysis}

This paper focuses on the study of a-posteriori identifiability to assess whether the parameters can be identified in a practical scenario with real experimental data. 

Practical identifiability tests considered in this study are based on the analysis of local sensitivity and correlation matrix since they are less expensive from the computational point of view compared to alternative methods such as Markov Chain Monte
Carlo, and their application has been validated in several works available in the literature \citep{wieland_structural_2021}.

\subsubsection{Local sensitivity analysis}

This analysis is \textit{local} because it is performed around a nominal value for parameters $\bm{\hat{\theta}}$., which can be estimated from preliminary data.
To construct the dynamic sensitivity matrix, $s_{ij} (t_k )$  the sensitivity of the i-th response $y_i$ to the $j$-th parameter 
$\hat{\theta}_j$ at the $k$-th sampling time $t_k$  is calculated as 
\begin{equation}
\label{eq:sens}
    s_{ij} (t_k )=\frac{\partial y_i (t_k)}{\partial \hat{\theta}_j }
\end{equation}
So that the dynamic sensitivity matrix $\bm{S}$ is obtained
\begin{equation}\label{eq:sens-matrix}
    \bm{S}_{N_y N_{sp} \times N_\theta } = \begin{bmatrix} 
    s_{11}(t_1) & \dots  & s_{1 N_\theta}(t_1)\\
    \vdots & \ddots & \vdots\\
    s_{11}(N_{sp}) & \dots  & s_{1 N_\theta}(t_{N_{sp}})\\     \vdots & \ddots & \vdots\\   
    s_{N_y1}(t_1) & \dots  & s_{N_y N_\theta}(t_1)\\
    \vdots & \ddots & \vdots\\
    s_{N_y1}(N_{sp}) & \dots  & s_{N_y N_\theta}(t_{N_{sp}})  
    \end{bmatrix}
\end{equation}
where $N_{sp}$ is the number of sampling points and $\bm{t}$ the  $N_{sp}$-dimensional vector of sampling times.

The dynamic sensitivity is evaluated over the whole time domain and the profiles plotted. If two or more parameters give overlapping profiles, this is an indication of practical non-identifiability, i.e. the corresponding parameters have the same effect on the system response and are correlated.

\subsubsection{Correlation matrix method} \label{sec:corr-matrix-method}

The correlation matrix approach is used to evaluate the identifiability of parameters, which relies on the matrix of sensitivities $\bm{S}$ computed in \equatref{eq:sens-matrix}. Given a preliminary estimate of the model parameters $\bm{\hat{\theta}}$, the matrix $\bm{S}$ is combined with the variance-covariance matrix of the measurements $\bm{\Sigma}_y$, i.e. a diagonal matrix with the observed variance from experimental replicates on the main diagonal, to calculate the Fisher information matrix $\bm{H}$ \citep{walter_identification_1997} as
\begin{equation} \label{eq:fim}
   \bm{H} = \bm{S}^T \bm{\Sigma}_y^{-1} \bm{S} + \bm{H}^0
\end{equation}
where $\bm{H}^0$ is the preliminary information on the parameters, which can be neglected if no prior information is available. 
The variance-covariance of the estimates $\bm{V}_{\theta} = \{ V_{\theta_{ij}}\}$ is approximated by the inverse of the observed $\bm{H}$ in the form
\begin{equation} \label{eq:vartheta}
    \bm{V}_{\theta} = \bm{H}^{-1} .
\end{equation}
The correlation matrix is then defined as $\bm{R} = \{r_{ij} \}$, where 
\begin{equation}\label{eq:corr_coef}
    r_{ij} = \frac{V_{\theta_{ij}}}{\sqrt{V_{\theta_{ii}} V_{\theta_{jj}} }} \quad \forall i,j = 1,\dots, N_\theta
\end{equation}

A correlation between parameters higher than 0.99 is a sign of practical non-identifiability, approaching a singular Fisher information matrix \citep{rodriguez-fernandez_novel_2006}. In this study, a conservative threshold of 0.95 is chosen as critical correlation.

\subsection{Data augmentation study} \label{sec:method-data-augmentation}
Two studies are performed to evaluate experimental conditions enabling statistically reliable model identification. Both data augmentation procedures are presented in the following subsections.

\subsubsection{Single additional sample}
The first data augmentation study is performed to assess the expected improvement in the statistics of parameter estimation, data fitting, and identifiability of model parameters deriving from the availability of additional experimental data, depending on the experimental design $\bm{\varphi}$ , i.e. the set of conditions at which the new data are collected.

The steps involved in this study are: 
\begin{enumerate}
    \item The experimental design vector $\bm{\varphi}$ is defined as \{$t_1,\dots,t_{N_{new}}$\}, where $N_{new}$ is the number of sampling times under assessment in the design space $\Phi$.
    \item The $N_{new}$  design variables $t_i$ are sampled using an equally spaced sampling of the design space defined by upper and lower bounds on the experimental sampling times.
    \item New experimental data are generated in silico from the model for each of the $N_{new}$ elements of the design vector $\bm{\varphi}$, and added to the original dataset $\psi_o$ to obtain $N_{new}$ augmented datasets:
    \begin{equation}
        \psi_i = \psi_o + \{\hat{\bm{y}}(\bm{x},\bm{u},\hat{\bm{\theta}}_0,t_i) + \bm{\varepsilon}_i\} \quad \forall  i=1,\dots,N_{new}
    \end{equation}
    where $\hat{\bm{\theta}}_0$ is the preliminary estimate of parameters, and the error $\bm{\varepsilon}_i~\in~\mathcal{N}(\bm{0},\bm{\sigma}_i^2)$ is obtained from a normal distribution with zero mean and variance $\bm{\sigma}_i^2$. The variance is obtained from a heteroscedastic model
    \begin{equation}\label{eq:heterosc-model}
        \bm{\sigma}_i^2 = \omega^2 (\hat{\bm{y}}^2)^{\gamma/2}
    \end{equation}
    where parameters $\omega$ and $\gamma$ are calibrated from the variance observed in the original experimental dataset.
    \item Perform parameter estimation for every $N_{new}$ augmented dataset $\psi_i$, and evaluate the statistics on the new estimates. The results are the new estimate $\bm{\theta}_i$, the covariance of the parameters $\bm{V}_{\bm{\theta}_i}$, the FIM $\bm{H}_i$, the t-values of the parameters $\bm{t}_{\bm{\theta}_i}$ (see Eq.~\eqref{eq:tvalue}) and the sum of squared residuals, i.e. the $\chi^2_i$ statistics, for all $i=1,\dots,N_{new}$.
    \begin{equation}\label{eq:tvalue}
        t_{\theta_j} = \frac{\hat{\theta}_j}{t(\frac{1+\alpha}{2})\sqrt{V_{\theta_{jj}}}} \quad \forall j=1,\dots,N_{\theta}
    \end{equation}
\end{enumerate}

In Eq.~\eqref{eq:tvalue}, the value $t(\frac{1+\alpha}{2})$ is obtained from a Student's distribution with $\dim(\psi) - N_{\theta}$ degrees of freedom and significance $\frac{1+\alpha}{2}$. The t-values of the parameters calculated as in Eq.~\eqref{eq:tvalue} are compared to a t-reference value $t(\alpha)$ given the significance level $\alpha$. 

This first data augmentation study is performed to understand under which conditions, i.e. sampling time, an additional experiment should be conducted so that the new data will carry more information in the modeling process.\\

\subsubsection{Multiple additional samples}
The second data augmentation study is performed to verify how many additional data are required to solve parameter identifiability issues, i.e. to estimate the full set of model parameters precisely. The procedure for this second study is the following:
\begin{enumerate}
    \item Select and fix the design space $\Phi$ for experimental design variables $\bm{\varphi}$ starting from the results of the previous study. 
    \item Select $n_{sp}$ sampling points, i.e. $\bm{\varphi}_{n_{sp}}=\{t_1,\dots,t_{n_{sp}}\}$, uniformly distributed in the design space $\Phi$.
    \item Generate new data $\psi_{new}$ in silico for each sampling point in $\bm{\varphi}_{n_{sp}}$,
    \begin{equation}
        \psi_{new} = \{\hat{\bm{y}}(\bm{x},\bm{u},\hat{\bm{\theta}}_0,t_i) + \bm{\varepsilon}_i\ | t_i\in \bm{\varphi}_{n_{sp}} \quad \forall i = 1,\dots,n_{sp}\}
    \end{equation}
    where the noise term $\bm{\varepsilon}_i$ is modeled as in Eq.~\eqref{eq:heterosc-model}.
    \item Add the new data $\psi_{new}$ to the original dataset available $\psi_o$ to obtain the augmented dataset $\psi_A$.
    \begin{equation}
        \psi_A = \psi_o + \psi_{new}
    \end{equation}
    \item Perform parameter estimation and evaluate the statistics, i.e. t-test on the estimates.
    \item Increase the number of sampling points and iterate the procedure from point 2., until the maximum budget is reached.
\end{enumerate}

The following section will present the case study on which the modeling framework is applied. i.e. foliar uptake of pesticides, in particular focusing on the models considered to describe the system behavior. 

\section{Foliar uptake models}
\label{sec:models}

In this paper different candidate models are compared for the description of pesticide uptake through the leaves. In particular the two models included in this study are $i$) a diffusion-based model \citep{sangoi_fosbe_2024}, and $ii$) a compartmental model \citep{sangoi_escape_2024}.


\subsection{Compartmental model} \label{sec:comp-model}
\begin{figure}
    \centering
    \includegraphics[width=0.6\linewidth]{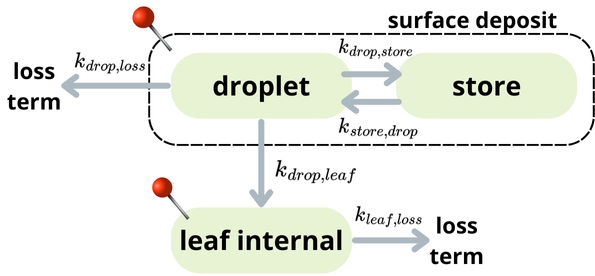}
    \caption{Graphical representation of the compartmental model for leaf uptake. The pins indicate the observed states, the compartments represent the system states and the arrows the transfer rates between the compartments (model parameters).}
    \label{fig:comp-model}
\end{figure}

The compartmental model included in this study to describe the foliar uptake process is presented graphically in Figure \ref{fig:comp-model}, where the pins indicate the observed states in the system. The compartments included in the model formulation are the following: droplet, store, leaf internal, and surface deposit. The \textit{droplet} is formulated product deposited on the leaf surface, the \textit{store} compartment contains the active ingredient (AI) crystallized on the surface and not available for uptake, while the \textit{surface deposit} is the sum of the two compartments, which corresponds to the system state observed experimentally. As for the leaf internal, a single compartment is considered because a more detailed division in the different layers that constitute the leaf would lead to a-priori non-identifiability issues when lacking of experimental observations of the different layers, as observed in a previous study \citep{sangoi_escape_2024}. Moreover, this choice ensure consistency with the diffusion-based model included in this study to have a fair comparison.

In Figure \ref{fig:comp-model}, the arrows indicate where the mass transfer between the compartments takes place. Mass transfer between the droplet and store compartments is associated to crystallization/solubility processes on the leaf surface \citep{burkhardt_stomatal_2012}, while the loss term from the droplet takes into account the AI lost, i.e. not available for uptake, due to volatility and/or photo-instability \citep{bronzato_measuring_2023}.

The AI in solution in the droplet is available for uptake in the leaf internal compartment ($k_{drop,leaf}$), and then the loss term $k_{leaf,loss}$ takes into account AI consumption inside the leaf due to metabolism or chemical-instabilities which reduce the amount of AI available in the leaf with time.

The mathematical expression of the compartmental model is reported in Equation \eqref{eq:compartmental}. The dynamic model is a system of ODEs, which describes the evolution in time of the AI mass in the different compartments. 

\begin{equation}\label{eq:compartmental}
    \frac{dm_i}{dt} = \sum_{j\neq i} (k_{ji}m_j - k_{ij} m_i)   
\end{equation}
In Equation \eqref{eq:compartmental}, $m_i$ is the mass of AI in compartment $i$, and $k_{ij}$ the transfer rate of AI from compartment $i$ to compartment $j$. The $m_i(t)$ values are normalized with respect to $m_{deposit}(t=0)$, therefore transfer rates $k_{ij}$ are expressed in units of $\text{min}^{-1}$. 

 With respect to the generic formulation presented in Equation \eqref{eq:generalmodel}, the vector of state variables is $\bm{x}:=\{m_i\}$, and the vector of model parameters is $\bm{\theta}:=\{k_{ij}\}$. No inputs $\bm{u}$ are present in the system. The vector of observable outputs is $\bm{y}:=\{ m_{deposit}, m_{leaf}\}$, where $m_{deposit}=m_{droplet}+m_{store}$. The experimental design vector is defined by the sampling times $\bm{\varphi}=\{t_1,\dots,t_{N_{sp}}\}$.

\subsection{Diffusion-based model}

\begin{figure}
    \centering
    \includegraphics[width=0.8\linewidth]{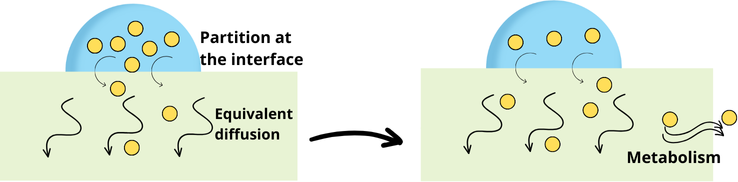}
    \caption{Graphical representation of the diffusion-based model for leaf uptake: system geometry and physical processes included.}
    \label{fig:diff-model}
\end{figure}

A second model considered in this study is a diffusion-based model. The geometry of the system and the physical phenomena included in its mathematical formulation are presented graphically in Figure \ref{fig:diff-model}. 
Previous studies in the literature \citep{schreiber_review_2006} suggest that the transport mechanism of pesticides through the cuticle, i.e. the external layer in the leaf structure protecting the cellular tissue from the external environment, can be assumed as diffusion. Although, separating the cuticle from the rest of the leaf is extremely complex and time consuming, therefore measuring the uptake in the cuticle  and leaf tissue separately is not an activity typically performed in routine biokinetic experimental procedures of foliar uptake of AIs. Since the purpose of this project is to validate a model that can be combined with the \textit{in vitro} and \textit{in vivo} experimental campaigns for the development of new biocides, in this diffusion-based model it is assumed that the \textit{leaf internal} is a homogeneous structure where \textit{equivalent diffusion} takes place, in the same way that only a single \textit{leaf internal} compartment is included in the compartmental model presented in Section \ref{sec:comp-model}.

The system geometry is then divided in two regions, as depicted in Figure~\ref{fig:diff-model}: the deposit on the surface and the leaf internal. The physical phenomena included in the model are: equilibrium at the interface between deposit and leaf, diffusion of AI through the leaf, and consumption of AI in the leaf due to metabolism.

The following equations are included in the general mathematical formulation developed to describe the dynamics of AI uptake from the deposit to the leaf internal region.

\begin{equation}\label{eq:diffmod-mdep}
     m_{dep}(t) = C_{dep}(t) \cdot V_{dep}(t)
\end{equation}

\begin{equation}\label{eq:diffmod-vdep}
    \frac{d V_{dep}(t)}{dt} = - K_{evap}\cdot f(V_{dep})
\end{equation}

subject to
\begin{align}\label{eq:diffmod-fv}
f(V_{dep}) =
    \begin{cases} 
        1 & \text{if } V_{dep} > V_{min}, \\
        0 & \text{if } V_{dep} \leq V_{min}.
    \end{cases}
\end{align}

\begin{equation}\label{eq:diffmod-cdep}
\frac{dC_{dep}}{dt} = - \frac{f(V_{dep})}{V_{dep}} \frac{dV_{dep}}{dt} C_{dep} - K_{loss} C_{dep}
\end{equation}

\begin{equation}\label{eq:diffmod-kdl}
K_{DL}= \frac{C_{leaf}(0,t)}{C_{dep} (t)}
\end{equation}

\begin{equation}\label{eq:diffmod-cleaf}
\frac{\partial C_{leaf}(z,t)}{\partial t} = D_{eq} \frac{\partial ^2 C_{leaf}(z,t)}{\partial z^2} -K_{met} C_{leaf} (z,t) - K_{trans} 
\end{equation}

\begin{equation}\label{eq:diffmod-cleaftot}
    C_{leaf,tot}(t) = \frac{1}{L_{leaf}} \int_{z=0}^{L_{leaf}} C_{leaf}(z,t)dz
\end{equation}

\begin{equation}\label{eq:diffmod-mleaf}
    m_{leaf}(t) = C_{leaf,tot}(t) \cdot V_{leaf}
\end{equation}

In the equations reported above, the state variables are the mass of AI in the deposit $m_{dep}$, the concentration of AI in the deposit $C_{dep}$, the concentration of AI inside the leaf discretized in space $C_{leaf}(z,t)$, the total concentration of AI inside the leaf $C_{leaf,tot}$, the mass of AI inside the leaf $m_{leaf}$. 
The parameters in the model are the evaporation rate $K_{evap}$, the accounting for losses from the deposit $K_{loss}$, the partition coefficient for AI between deposit and leaf $K_{DL}$, the diffusion coefficient inside the leaf $D_{eq}$, the metabolism rate $K_{met}$, the translocation to other parts of the leaf/plant $K_{trans}$.

Similarly to the compartmental model, also for the diffusion model the state variables are normalized with respect to the initial amount of AI in the deposit. 

Equations \eqref{eq:diffmod-mdep}-\eqref{eq:diffmod-cdep} describe the dynamics in the deposit on the leaf surface. Equation~\eqref{eq:diffmod-kdl} relates the concentration of AI in the deposit to the concentration in the leaf at the interface, assuming equilibrium conditions at all times within a boundary layer at the interface between leaf and droplet. In Equation~\eqref{eq:diffmod-kdl}, $C_{leaf} (0,t)$ indicates the concentration of AI at the interface on the leaf side at time $t$.
Equations~\eqref{eq:diffmod-cleaf}-\eqref{eq:diffmod-mleaf} describe the dynamics in the leaf tissue. The leaf internal region is modeled as a homogeneous structure where diffusion and metabolic consumption take place uniformly throughout the spatial domain.

\section{Results and discussion}
\label{sec:results}

The analyses on the compartmental and diffusion-based models presented in Section \ref{sec:models} are conducted starting from experimental data of foliar uptake provided by Syngenta. The same procedure and analyses are conducted on two different datasets, named as TR-1 (treatment 1) and TR-2 (treatment 2). The experimental data represent biokinetic experiments of foliar uptake, where two quantities are measured for each sampling time: the amount of active ingredient (AI) left on the leaf surface (i.e. \textit{deposit}), and the amount of AI inside the leaf (i.e. \textit{leaf extract}). For each dataset 8 sampling times from the application of the product on the leaf are considered, ranging from 0 to 360 minutes. 

The two dataset differ for the treatment depending if the leaves are subject to the solar radiation or not, while the combination of AI, formulation and crop are the same for the two treatments considered here.

The results of the analyses are presented with this structure: i) parameter estimation, data fitting and statistical tests (Section~\ref{sec:results-PE-fitting}), ii) model identifiability tests (Section~\ref{sec:results-identifiability}), iii) data augmentation study (Section~\ref{sec:results-data}).

\subsection{Parameter estimation, data fitting and statistical tests} \label{sec:results-PE-fitting}

The parameter estimation results (Table \ref{tab:estimate}) 
are given in terms of estimated values and 95\% confidence intervals obtained after a log-likelihood parameter estimation has been carried out, for both the datasets TR-1 and TR-2. For the diffusion-based model the estimated parameters are $K_{DL}$, $D_{eq}$ and $K_{met}$, assuming that the other parameters are negligible and setting their value to 0. 

For both models it is noted that the estimated values of the corresponding model parameters do not change significantly between TR-1 and TR-2, especially when considering the parametric uncertainty. A difference is noted in the compartmental model for the parameter $k_{drop,loss}$, where the absence of UV radiation in TR-2 leads to a lower value of this parameter, however this difference is still within the 95\% confidence intervals.
It must be highlighted that the uncertainty on the estimates is large, in particular for parameters $D_{eq}$ and $K_{DL}$ in the diffusion based model.

\begin{figure}[htp]
    \centering
    \begin{subfigure}[b]{0.49\linewidth}
        \includegraphics[width=\linewidth]{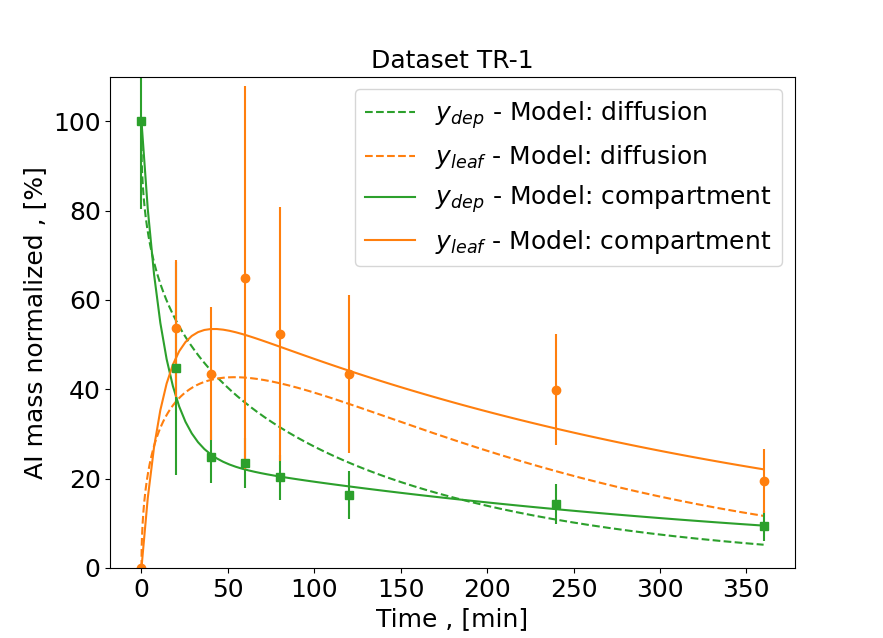}
        \subcaption{}
        \label{fig:fitting-tr1}
    \end{subfigure}
    \hfill
    \begin{subfigure}[b]{0.49\linewidth}
        \includegraphics[width=\linewidth]{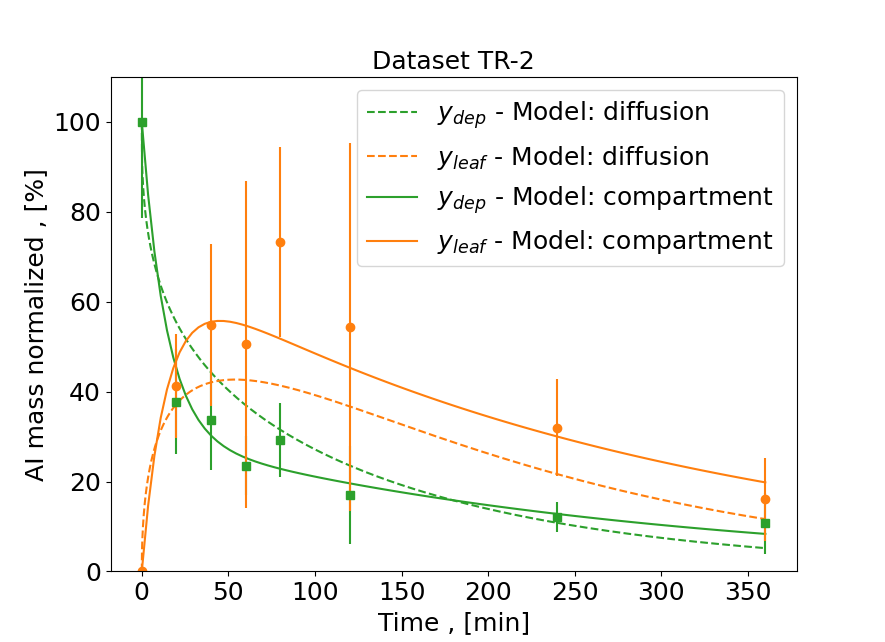}
        \subcaption{}
        \label{fig:fitting-tr2}
    \end{subfigure}
    \caption{Fitting of experimental data with the compartmental model (solid line) and diffusion-based model (dashed line). Results for (a) dataset TR-1 and (b) dataset TR-2. Green colour refers to measurements of the deposit on the leaf surface, orange colour to measurement in the leaf tissue.}
    \label{fig:fitting}
\end{figure}

The predicted profiles after parameter estimation with the compartmental and diffusion-based models are shown in \figref{fig:fitting}, along with the experimental data used to calibrate the model parameters. The error bars show the 95\% uncertainty region on the experimental data. It is observed that the compartmental model captures the data better than the diffusion-based model.

\begin{table}[htp]
    \centering
    \caption{Parameter estimates and 95\% confidence intervals for the compartmental and diffusion-based models obtained for both the datasets (TR-1 and TR-2).}
    \begin{tabular}{llllll}
        \toprule
            \textbf{Model} & \textbf{Dataset} & \textbf{Parameter} & \textbf{Estimate} & \textbf{$\pm$ 95\% C.I.} & \textbf{Units}\\
        \toprule
        Compartmental & TR-1 & $k_{drop,store}$  & 0.01938 & $\pm$ 0.01593 & 1/min\\
        Compartmental & TR-1 & $k_{store,drop}$  & 0.00363 & $\pm$ 0.00195 & 1/min\\
        Compartmental & TR-1 & $k_{drop,leaf}$   & 0.05004 & $\pm$ 0.01390 & 1/min\\
        Compartmental & TR-1 & $k_{drop,loss}$   & 0.01246 & $\pm$ 0.00779 & 1/min\\
        Compartmental & TR-1 &  $k_{leaf,loss}$  & 0.00382 & $\pm$ 0.00072 & 1/min\\
        \hline
        Compartmental & TR-2 &  $k_{drop,store}$  & 0.01863 & $\pm$ 0.01456 & 1/min\\
        Compartmental & TR-2 &  $k_{store,drop}$  & 0.00497 & $\pm$ 0.00312 & 1/min\\
        Compartmental & TR-2 &  $k_{drop,leaf}$   & 0.04578 & $\pm$ 0.01819 & 1/min\\
        Compartmental & TR-2 &  $k_{drop,loss}$   & 0.00455 & $\pm$ 0.01112 & 1/min\\
        Compartmental & TR-2 &  $k_{leaf,loss}$   & 0.00483 & $\pm$ 0.00176 & 1/min\\
        \hline
        Diffusion-based & TR-1 & $D_{eq}$  & 6.143 e-14 & $\pm$ 5.573 e-13 & m$^2$/s \\
        Diffusion-based & TR-1 & $K_{DL}$  & 3.941 e+01 & $\pm$ 1.767 e+02 & - \\
        Diffusion-based & TR-1 & $k_{met}$ & 4.525 e-02 & $\pm$ 9.365 e-03 & 1/s \\
        \hline        
        Diffusion-based & TR-2 & $D_{eq}$  & 4.587 e-14 & $\pm$ 3.608 e-13 & m$^2$/s \\
        Diffusion-based & TR-2 & $K_{DL}$  & 3.808 e+01 & $\pm$ 1.483 e+02 & - \\
        Diffusion-based & TR-2 & $k_{met}$ & 4.927 e-02 & $\pm$ 1.210 e-02 & 1/s \\
        \bottomrule
    \end{tabular}
    \label{tab:estimate}
\end{table}

\subsubsection{Statistical tests} \label{sec:results-stats}
The data fitting results are reported in Table~\ref{tab:fitting}. The sum of squared residuals (SSR) shows that the compartmental model has a better fitting than the diffusion model, since the SSR values obtained with the compartmental model are lower with both datasets. 
The table reports also the results of the $\chi^2$-test performed to evaluate the quality of data fitting. For the compartmental model the test is passed in both cases, while the fitting obtained with the diffusion-based model is not statistically acceptable for the dataset TR-1, since the SSR is higher than $\chi^2_{0.95}$, which means that the model is under-fitting the experimental data. Conversely, the diffusion-based model provides an adequate fitting when calibrated on TR-2 data.

\begin{table}[htp]
    \centering
    \caption{Quality of data fitting assessed comparing the sum of squared residuals (SSR) to the $\chi^2$-reference values at 0.05 and 0.95 significance.}
    \begin{tabular}{llllll}
        \toprule
            \textbf{Dataset} & \textbf{Model} & \textbf{SSR} & \textbf{$\chi^2_{0.05}$} & \textbf{$\chi^2_{0.95}$} & \textbf{Test result} \\
        \toprule
        TR-1  & Compartmental   & 6.558  & 4.575 & 19.675 & Passed \\
        TR-1  & Diffusion-based & 27.141 & 5.892 & 22.362 & Failed (under-fitting) \\
        \hline
        TR-2  & Compartmental   & 11.343 & 4.575 &  19.675 & Passed \\
        TR-2  & Diffusion-based & 17.620 & 5.892 & 22.362  & Passed \\
        \bottomrule
    \end{tabular}
    \label{tab:fitting}
\end{table}

\begin{figure}[htp]
    \centering
    \begin{subfigure}[b]{0.49\linewidth}
        \includegraphics[width=\linewidth]{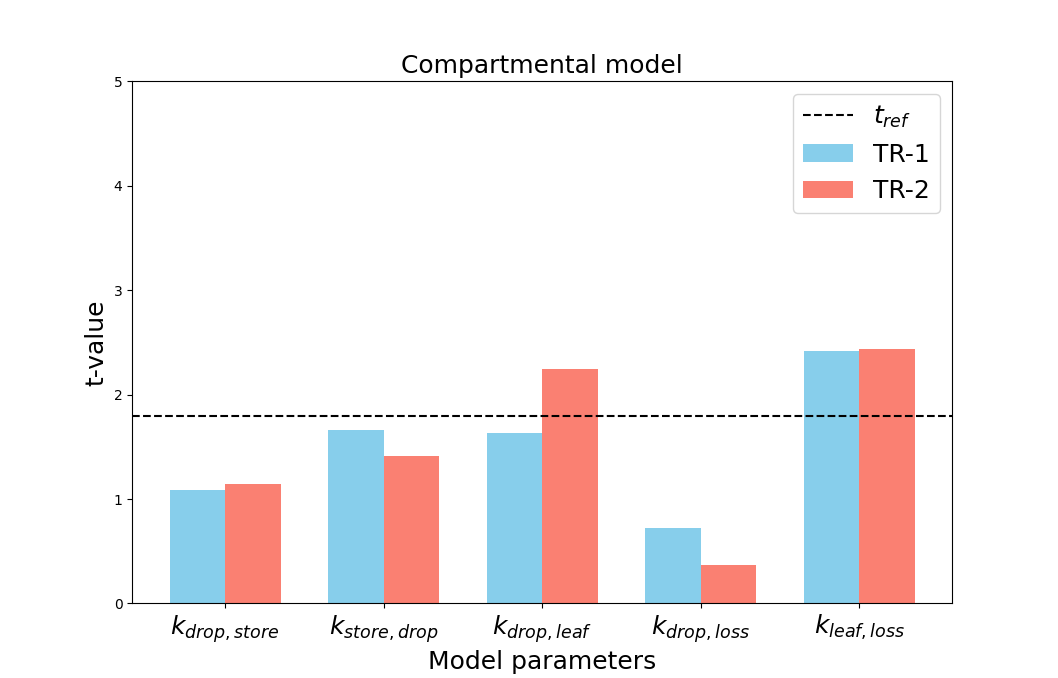}
        \subcaption{}
        \label{fig:ttest-comp}
    \end{subfigure}
    \hfill
    \begin{subfigure}[b]{0.49\linewidth}
        \includegraphics[width=\linewidth]{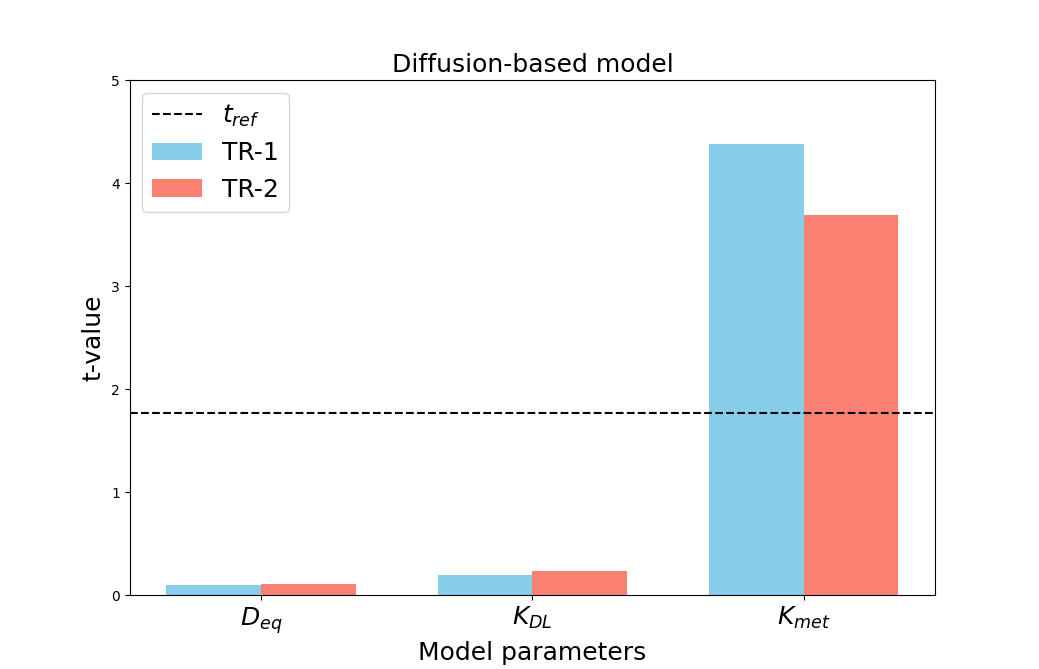}
        \subcaption{}
        \label{fig:ttest-diff}
    \end{subfigure}
    \caption{Bar chart with the values of the t-test statistics obtained on two datasets TR-1 and TR-2 for (a) the compartmental model and (b) the diffusion-based model parameters.}
    \label{fig:ttest}
\end{figure}

The statistical quality of the estimates is evaluated by means of a t-test, which results are shown in Figure~\ref{fig:ttest}. The compartmental model results (Fig.~\ref{fig:ttest-comp}) depict that the parameter $k_{leaf,loss}$, representing a consumption term inside the leaf, is estimated with a good confidence from both TR-1 and TR-2 datasets, being the t-values higher than the reference $t_{ref}$. For $k_{drop,leaf}$ the estimate is satisfactory only when the dataset TR-2 is used, while the t-values of all the other parameters are clearly lower than the reference. In particular, the loss term from the droplet $k_{drop,loss}$  is the most critical parameter to estimate, as underlined by the least confidence in the estimate.

Similar results are obtained with the diffusion-based model (Fig.~\ref{fig:ttest-diff}). In this case, the parameter $K_{met}$ that describes the AI consumption inside the leaf provides a statistically precise estimation from both datasets, similarly to $k_{leaf,loss}$ in the compartmental model. On the other hand, parameters $D_{eq}$ and $K_{DL}$ are poorly estimated, with very low t-values compared to the reference, which is a result that goes along with the large variances observed in Table~\ref{tab:estimate}.

\subsection{Model identifiability} \label{sec:results-identifiability}
To answer the question whether the results on the estimates could be improved by designing new and more informative experiments, the framework presented in the methodology, Section \ref{sec:methods} (Fig.~\ref{fig:framework}), is followed. The first analysis conducted is to assess the identifiability of the model parameters by analyzing the dynamic sensitivity profiles, which indicate the impact of the parameters on the output variables in time, then followed by the correlation analysis between the parameters.

\subsubsection{Local sensitivity profiles}
\begin{figure}[htp]
    \centering
    \begin{subfigure}[b]{0.49\linewidth}
        \centering
        \includegraphics[width=\linewidth]{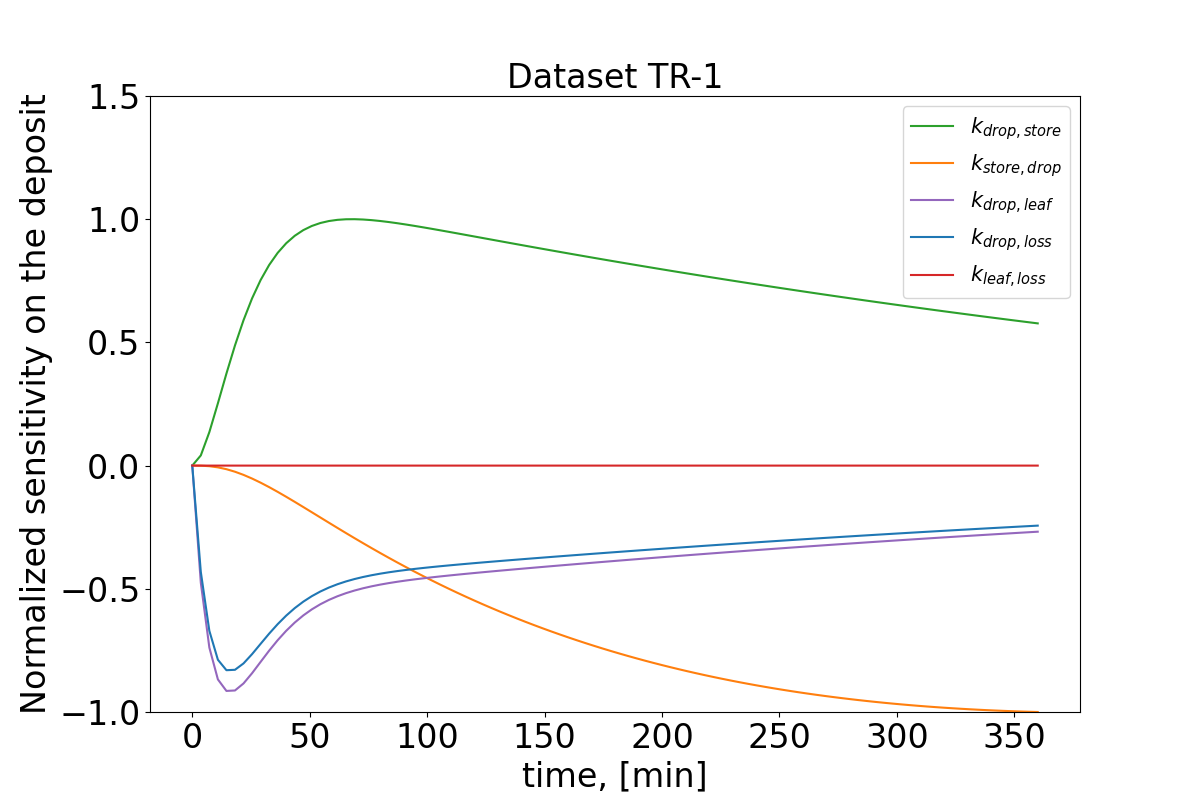}
        \caption{}
        \label{fig:sens-comp-TR1-dep}
    \end{subfigure}
    \hfill
    \begin{subfigure}[b]{0.49\linewidth}
        \centering
        \includegraphics[width=\linewidth]{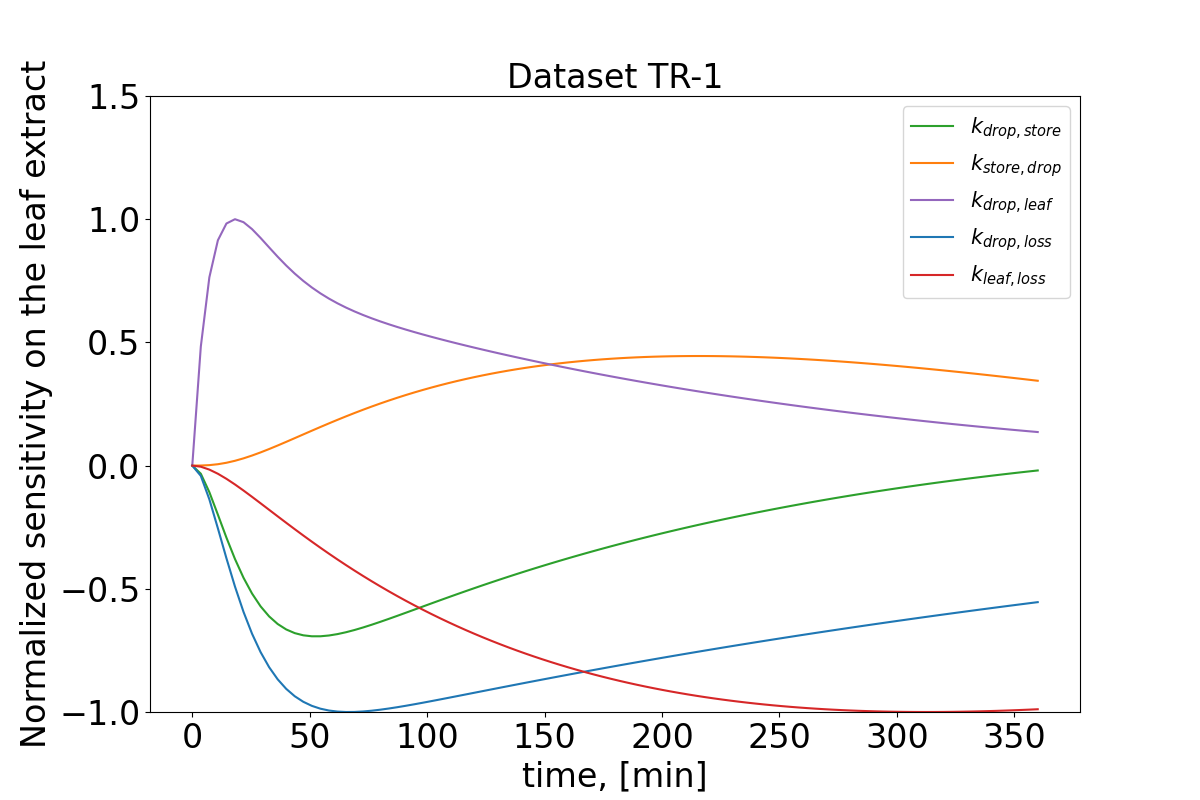}
        \caption{}
        \label{fig:sens-comp-TR1-leaf}
    \end{subfigure}
    \\
    \begin{subfigure}[b]{0.49\linewidth}
        \centering
        \includegraphics[width=\linewidth]{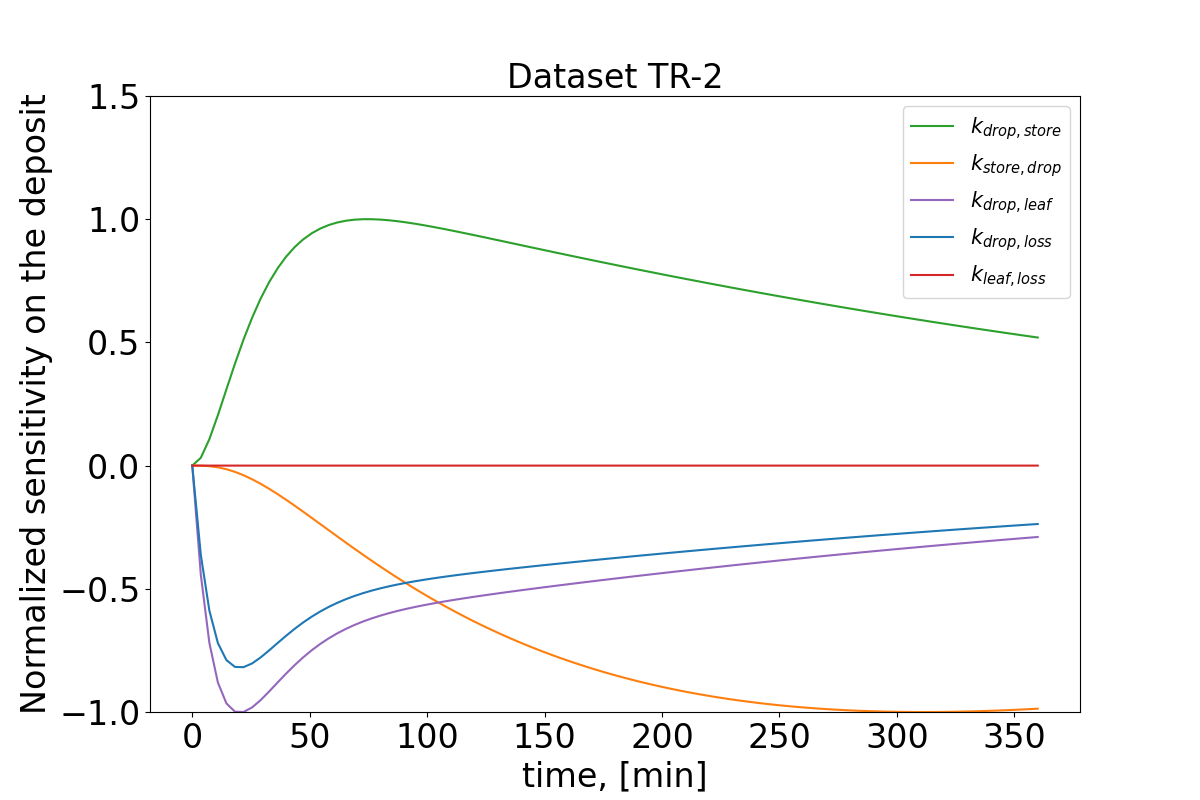}
        \caption{}
        \label{fig:sens-comp-TR2-dep}
    \end{subfigure}
    \hfill
    \begin{subfigure}[b]{0.49\linewidth}
        \centering
        \includegraphics[width=\linewidth]{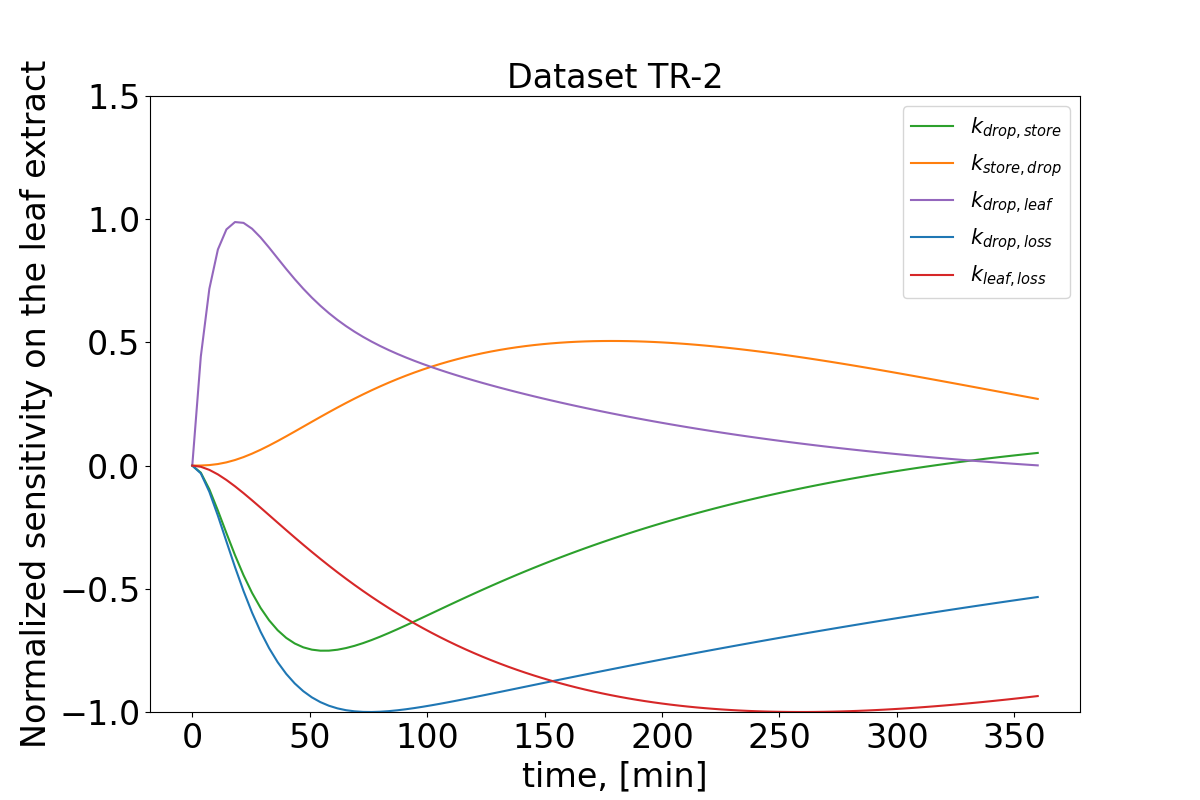}
        \caption{}
        \label{fig:sens-comp-TR2-leaf}
    \end{subfigure}
    \caption{Sensitivity profiles for the compartmental model parameters. (a) Output: deposit, dataset TR-1; (b) output: leaf extract, dataset TR-1; (c) output: deposit, dataset TR-2; (d) output: leaf extract, dataset TR-2.}
    \label{fig:sensitivity-comp}
\end{figure}

\begin{figure}[htp]
    \centering
    \begin{subfigure}[b]{0.49\linewidth}
        \centering
        \includegraphics[width=\linewidth]{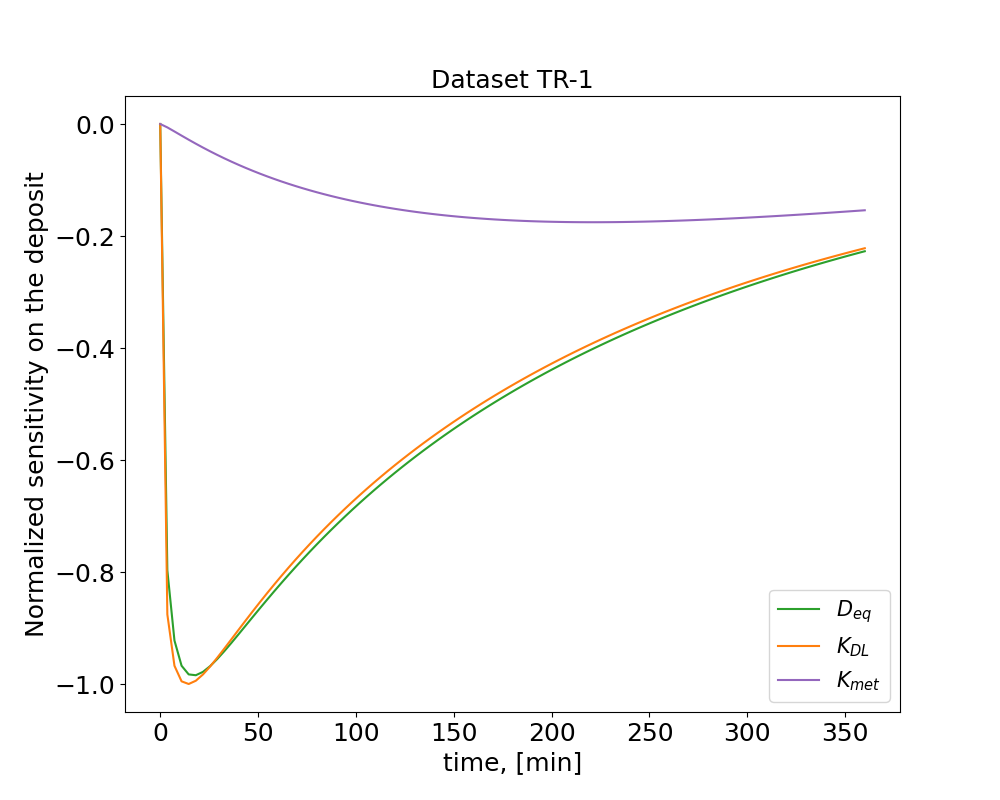}
        \caption{}
        \label{fig:sens-diff-TR1-dep}
    \end{subfigure}
    \hfill
    \begin{subfigure}[b]{0.49\linewidth}
        \centering
        \includegraphics[width=\linewidth]{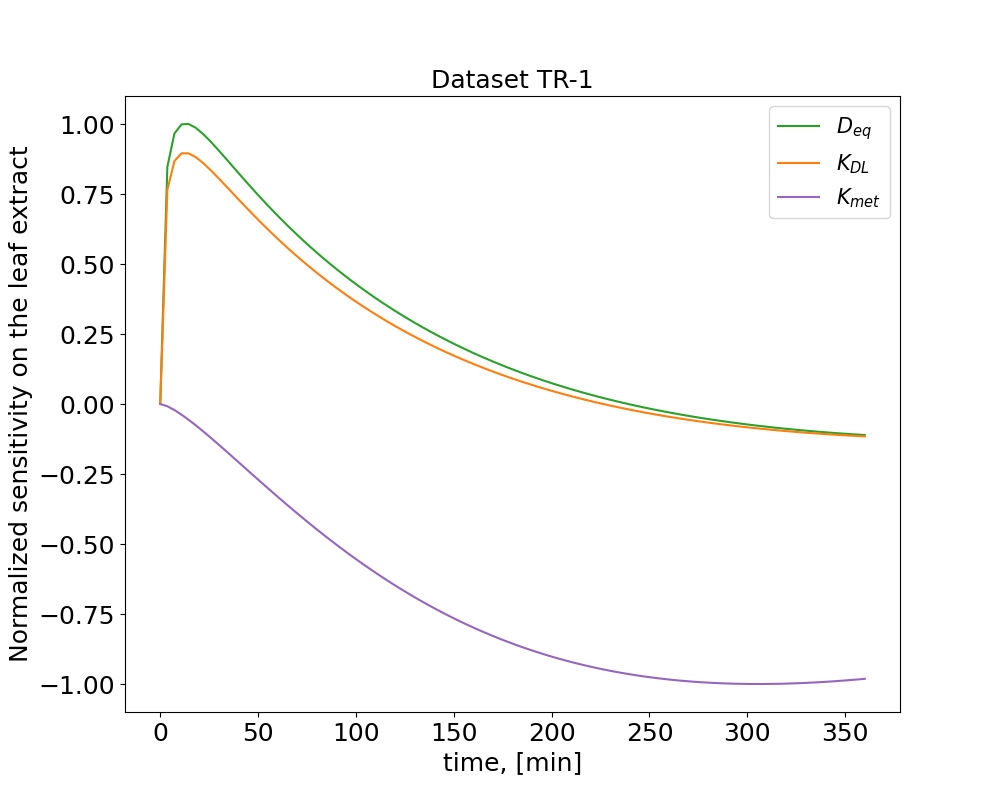}
        \caption{}
        \label{fig:sens-diff-TR1-leaf}
    \end{subfigure}
    \\
    \begin{subfigure}[b]{0.49\linewidth}
        \centering
        \includegraphics[width=\linewidth]{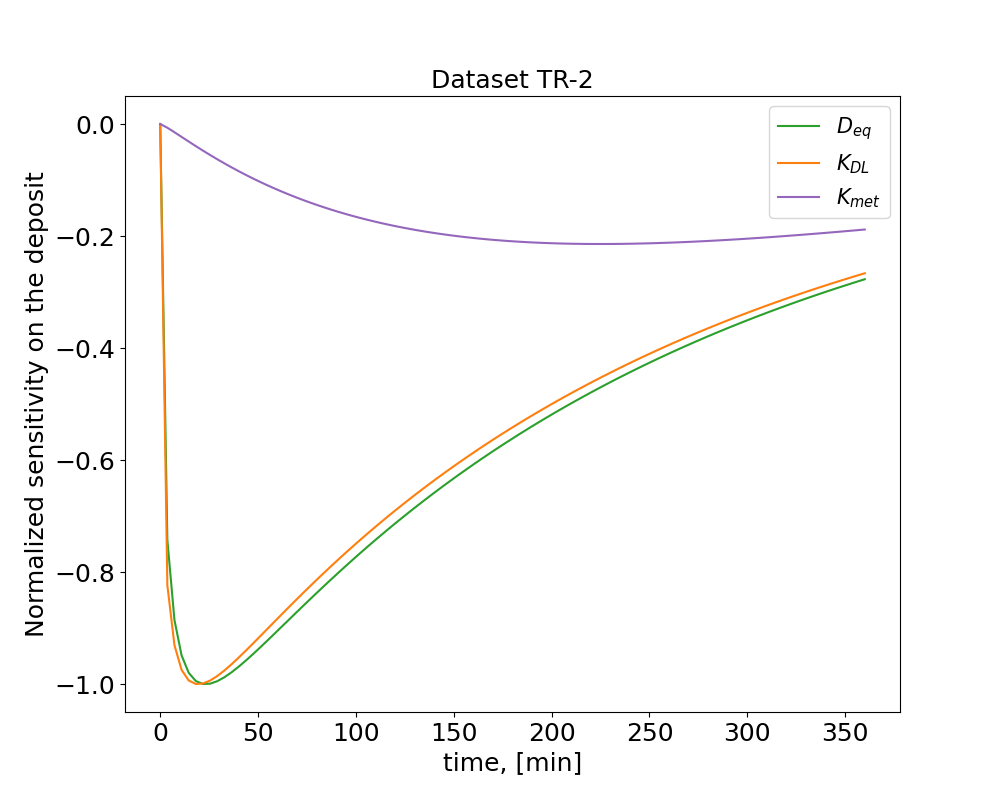}
        \caption{}
        \label{fig:sens-diff-TR2-dep}
    \end{subfigure}
    \hfill
    \begin{subfigure}[b]{0.49\linewidth}
        \centering
        \includegraphics[width=\linewidth]{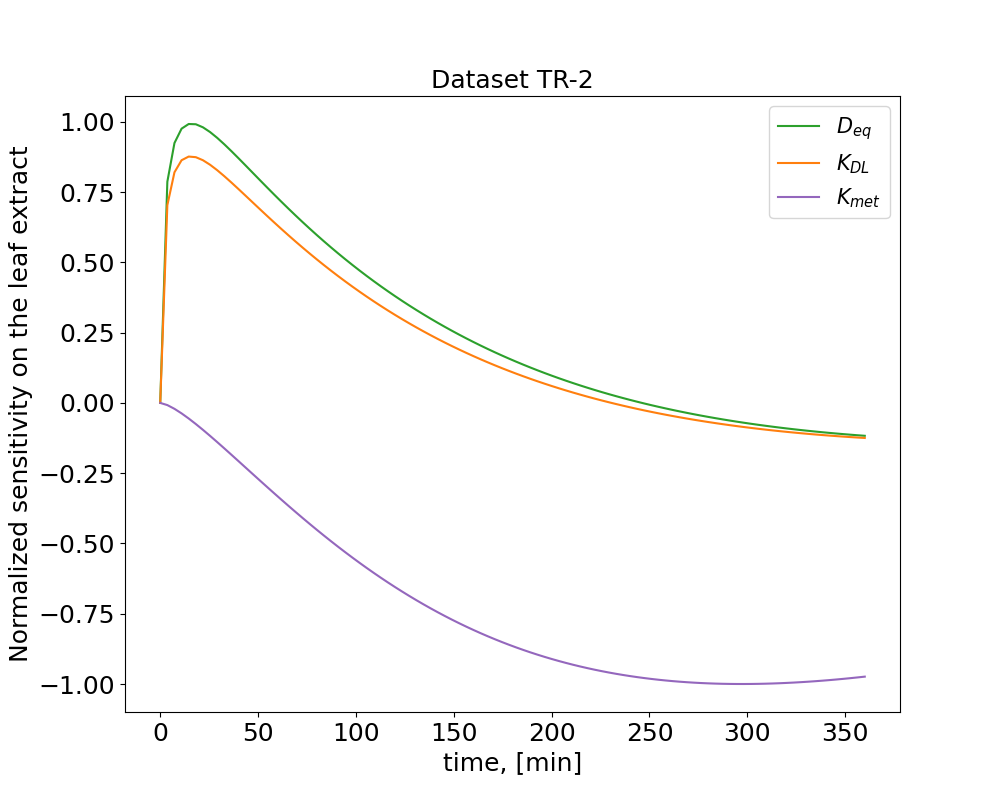}
        \caption{}
        \label{fig:sens-diff-TR2-leaf}
    \end{subfigure}
    \caption{Sensitivity profiles for the diffusion-based model parameters. (a) Output: deposit, dataset TR-1; (b) output: leaf extract, dataset TR-1; (c) output: deposit, dataset TR-2; (d) output: leaf extract, dataset TR-2.}
    \label{fig:sensitivity-diff}
\end{figure}

Results from the sensitivity analysis are shown in Figure~\ref{fig:sensitivity-comp} for the compartmental model and Figure~\ref{fig:sensitivity-diff} for the diffusion model. This is a local analysis performed around the preliminary estimate of the parameters, and for both models the results are reported starting from the estimate obtained with TR-1 and TR-2. It is observed that the profiles obtained for a given model with the two datasets are similar in their features, so only the profiles for TR-1 will be commented for the sake of conciseness. 
The values of sensitivity have been normalized for each parameter in the range between -1.0 and +1.0 to show which output measurement has the higher impact for a given parameter.

For the compartmental model it emerges that parameters $k_{drop,store}$ and $k_{store,drop}$ are mostly linked to the measurement on the deposit (Fig.~\ref{fig:sens-comp-TR1-dep}), $k_{leaf,loss}$ depends uniquely on the leaf extract measurement (Fig.~\ref{fig:sens-comp-TR1-leaf}), while both outputs are sensitive to the remaining two parameters $k_{drop,leaf}$ and $k_{drop,loss}$. It must also be highlighted that the loss due to AI consumption in the leaf ($k_{leaf,loss}$) has the peak of sensitivity between 3 and 6 hours since the deposition of the droplet on the leaves, while impact on the measured outputs of the parameters $k_{drop,store}$, $k_{drop,leaf}$ and $k_{drop,loss}$ is stronger in the short time from the deposition (within 1 hour), when the initial uptake dynamics takes place.  

The sensitivity profiles obtained with the diffusion model are reported in Figure~\ref{fig:sensitivity-diff}. Also in this case it is observed that the metabolic rate is mostly linked to the leaf extract output (Fig.~\ref{fig:sens-diff-TR1-leaf}), and that its effect peaks between 3 and 6 hours after the deposition of the formulation on the leaves, similarly to what has been observed with $k_{leaf,loss}$ in the compartmental model. The other two parameters $K_{DL}$ and $D_{eq}$ impact both outputs equally, and their peak is observed in the first hour from the deposition of the droplets. It is important to note that the profiles for these two parameters are overlapping, which indicates that their relation to the observable states is identical, implying practical identifiability issues for $K_{DL}$ and $D_{eq}$.

\subsubsection{Parameter correlation analysis}
\begin{figure}[htp]
    \centering
    \begin{subfigure}[b]{0.49\linewidth}
        \centering
        \includegraphics[width=\linewidth]{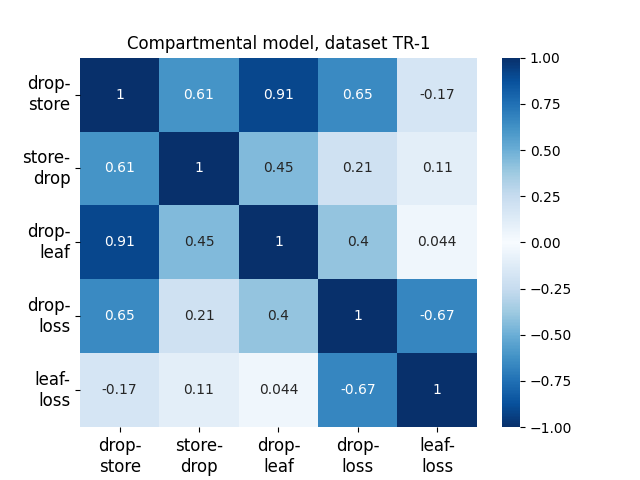}
        \caption{}
        \label{fig:corr-comp-TR1}
    \end{subfigure}
    \hfill
    \begin{subfigure}[b]{0.49\linewidth}
        \centering
        \includegraphics[width=\linewidth]{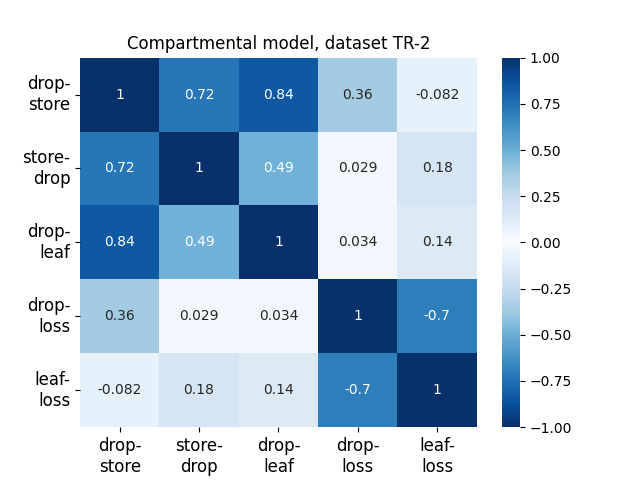}
        \caption{}
        \label{fig:corr-comp-TR2}
    \end{subfigure}
    \\
    \begin{subfigure}[b]{0.49\linewidth}
        \centering
        \includegraphics[width=\linewidth]{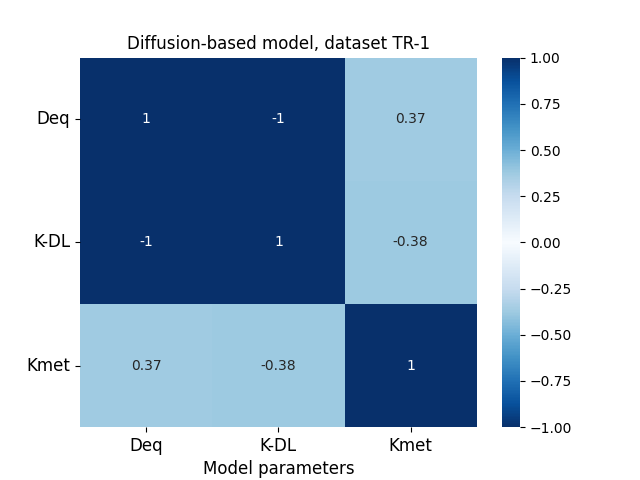}
        \caption{}
        \label{fig:corr-diff-TR1}
    \end{subfigure}
    \hfill
    \begin{subfigure}[b]{0.49\linewidth}
        \centering
        \includegraphics[width=\linewidth]{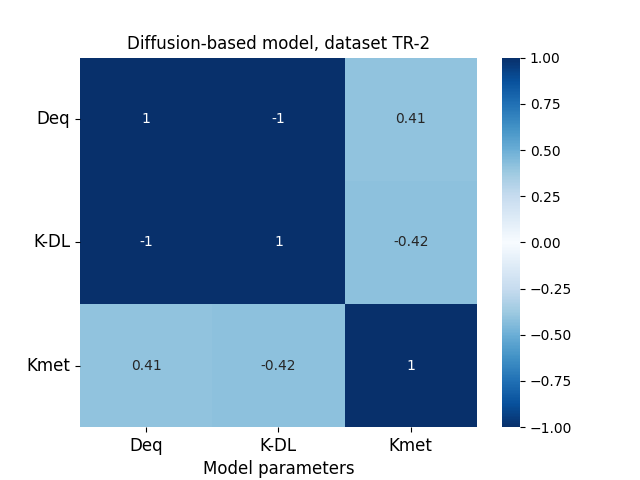}
        \caption{}
        \label{fig:corr-diff-TR2}
    \end{subfigure}
    \caption{Correlation matrix for the model parameters: (a) compartmental model, dataset TR-1; (b) compartmental model, dataset TR-2; (c) diffusion-based model, dataset TR-1; (a) diffusion-based model, dataset TR-2.}
    \label{fig:corr-matrix}
\end{figure}

Practical identifiability issues for $K_{DL}$ and $D_{eq}$ in the diffusion model, emerged from the dynamic sensitivity profiles, are further tested by studying the correlation between model parameters, following the methodology presented in Section~\ref{sec:methods}.
The results of this analysis are reported in Figure~\ref{fig:corr-matrix} by means of correlation matrices, respectively for compartmental model (Fig.~\ref{fig:corr-comp-TR1}, \ref{fig:corr-comp-TR2}) and diffusion model (Fig.~\ref{fig:corr-diff-TR1}, \ref{fig:corr-diff-TR2}). Similar results are obtained when conducting the analysis starting with different sets of preliminary data, i.e. TR-1 and TR-2, for a given model among the candidates. 

For the compartmental model, the matrix in Figure~\ref{fig:corr-comp-TR1}-\ref{fig:corr-comp-TR2} show a block diagonal trend where $k_{leaf,loss}$ is slightly correlated to $k_{drop,loss}$, while the other three parameters $k_{drop,store}$, $k_{store,drop}$ and $k_{drop,leaf}$ are moderately inter-correlated. However, the maximum correlation index observed between pairs of parameters in the compartmental model is 0.91 (for $k_{drop,store}$ and $k_{drop,leaf}$), which is below the conservative threshold of 0.95 defined in the methodology. Therefore, the analysis tells that practical identifiability is not an issue for the compartmental model parameters, whose estimates could then be improved by providing additional data from properly designed foliar uptake experiments.

The correlation matrix of the diffusion model parameters in Figure~\ref{fig:corr-diff-TR1}-\ref{fig:corr-diff-TR2} confirms the practical identifiability issue between $D_{eq}$ and $K_{DL}$. In both cases, using the data of TR-1 and TR-2, the correlation coefficient between these two parameters is -1, meaning that these parameters are totally anti-correlated and their effect cannot be decoupled with the given observed measurements. Conversely, the parameter $K_{met}$ is not strongly correlated to the others, which also explains the good estimate obtained for it.

\subsubsection{Likelihood function profiles}

\begin{figure}[htp]
    \centering
    \begin{subfigure}[b]{0.49\linewidth}
        \centering
        \includegraphics[width=\linewidth]{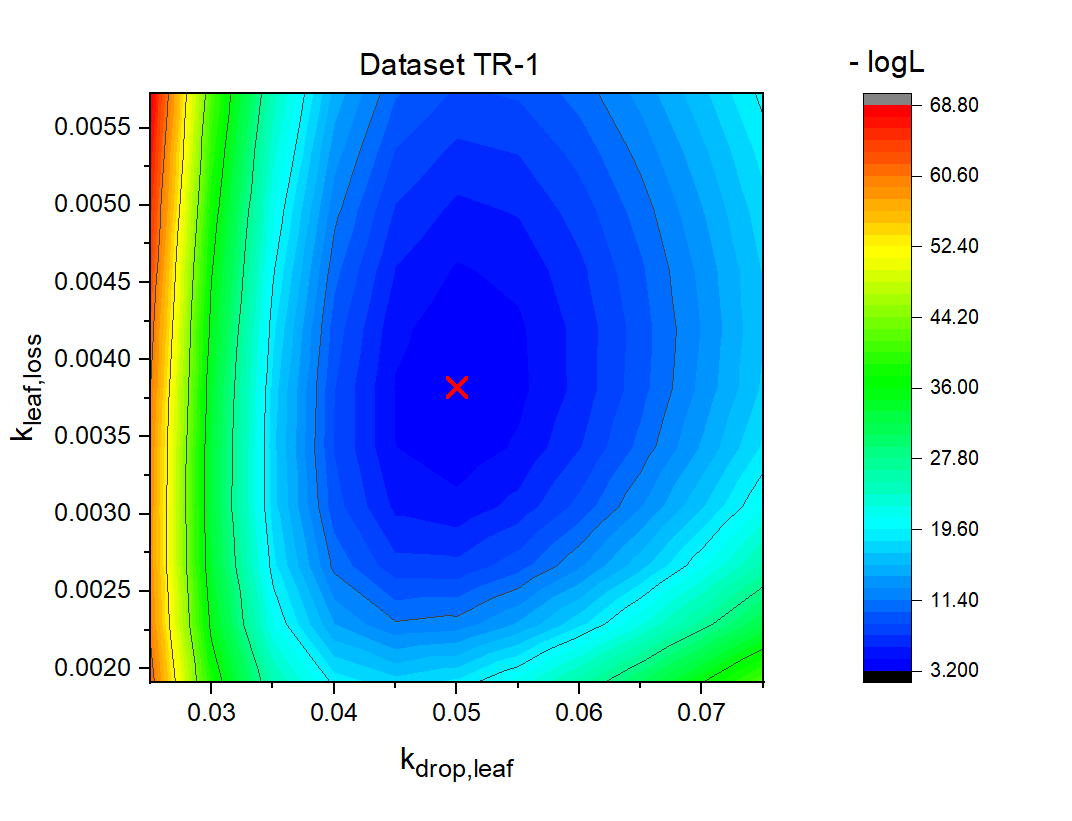}
        \caption{}
    \end{subfigure}
    \hfill
    \begin{subfigure}[b]{0.49\linewidth}
        \centering
        \includegraphics[width=\linewidth]{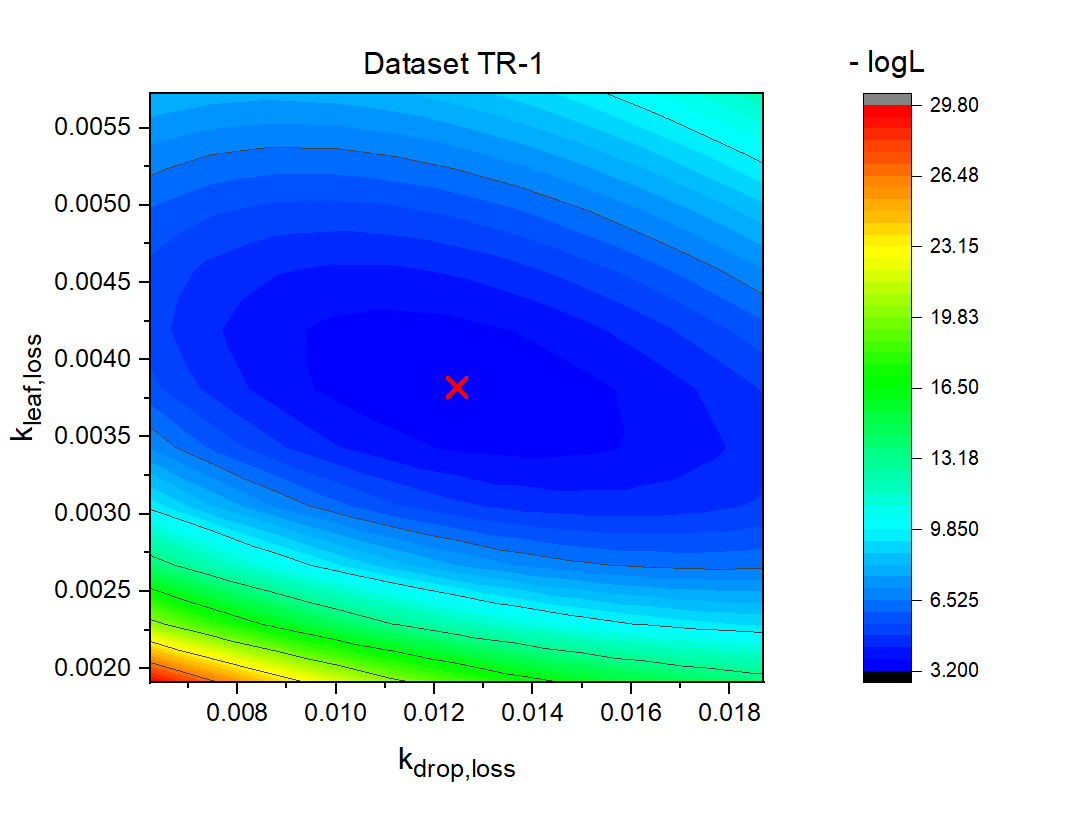}
        \caption{}
    \end{subfigure}
    \\
    \begin{subfigure}[b]{0.49\linewidth}
        \centering
        \includegraphics[width=\linewidth]{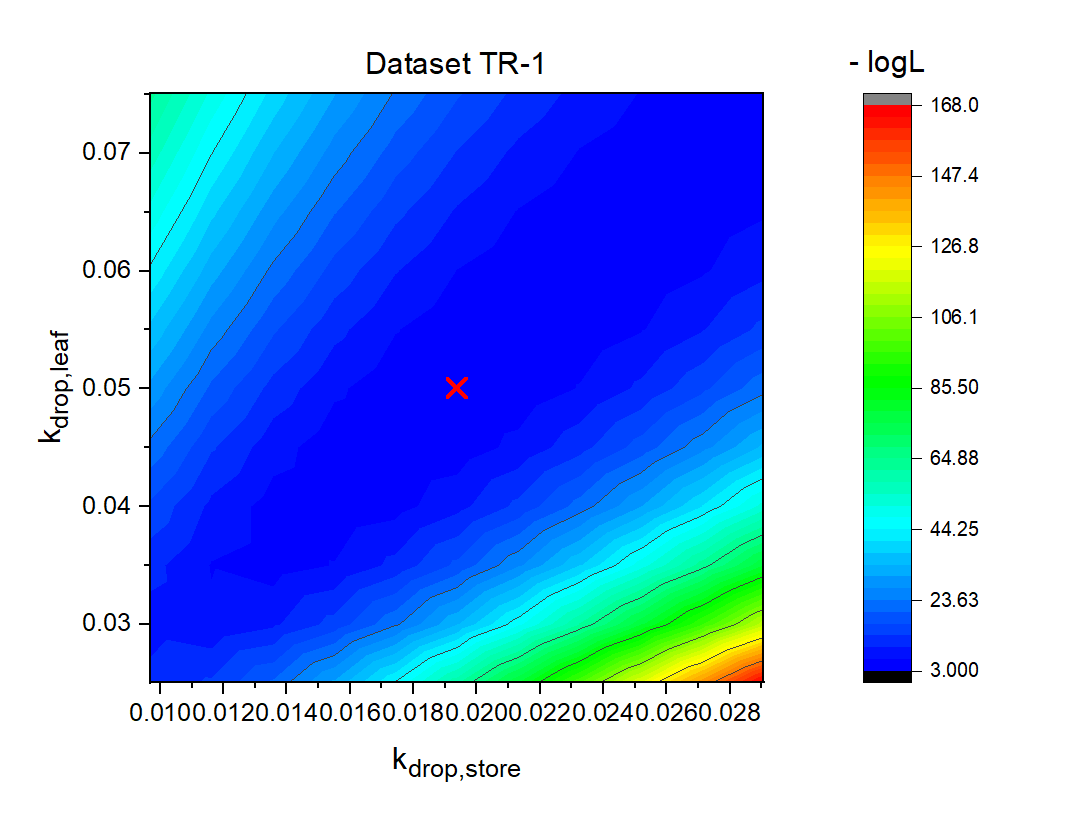}
        \caption{}
        \label{fig:likelihood-contours-compartmental-tr1-c}
    \end{subfigure}
    \hfill
    \begin{subfigure}[b]{0.49\linewidth}
        \centering
        \includegraphics[width=\linewidth]{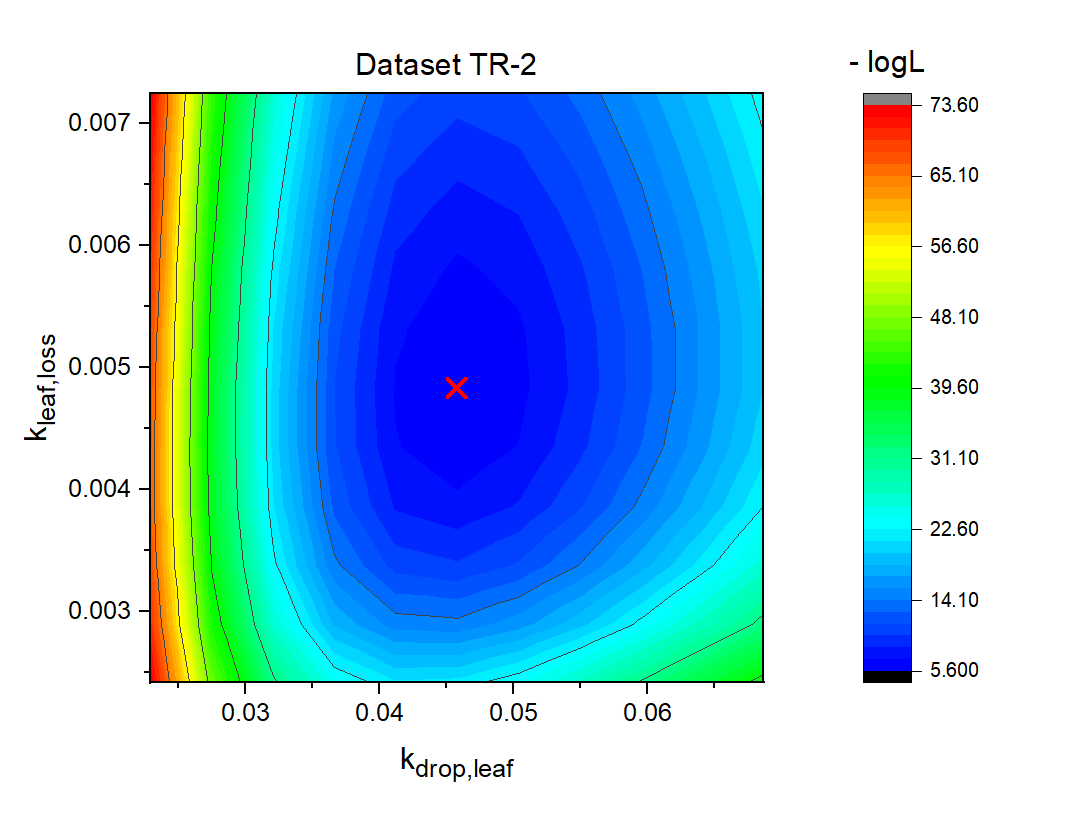}
        \caption{}
    \end{subfigure}
    \\
    \begin{subfigure}[b]{0.49\linewidth}
        \centering
        \includegraphics[width=\linewidth]{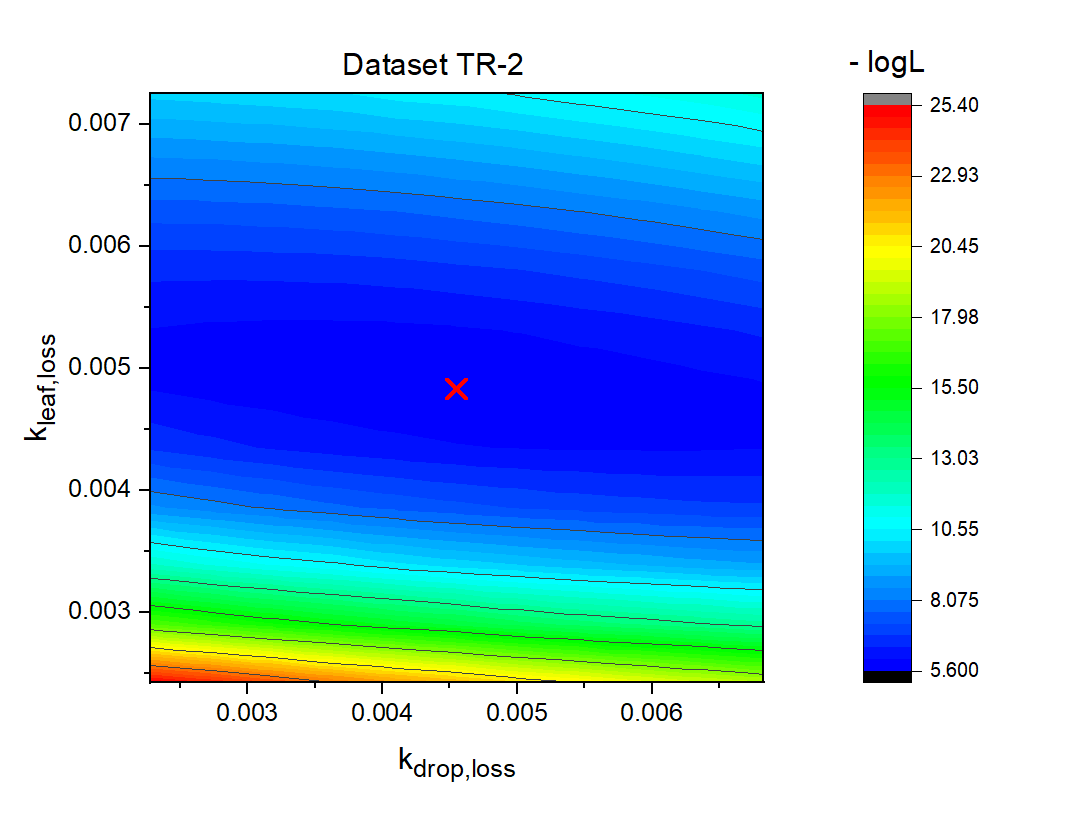}
        \caption{}
    \end{subfigure}
    \hfill
    \begin{subfigure}[b]{0.49\linewidth}
        \centering
        \includegraphics[width=\linewidth]{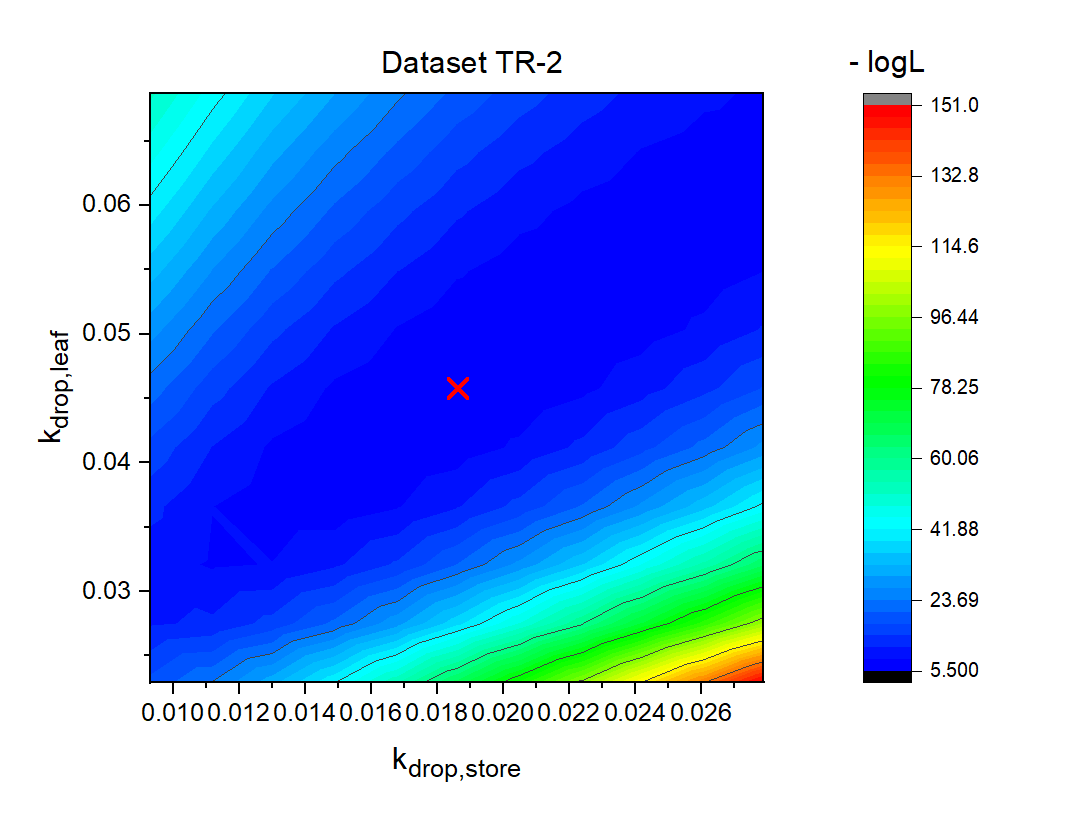}
        \caption{}
        \label{fig:likelihood-contours-compartmental-tr2-c}
    \end{subfigure}    
    \caption{Contour plots showing the likelihood function dependence on pairs of parameters for the compartmental model. The red cross is the parameter estimate. Dataset: (a), (b), (c)  TR-1; (d), (e), (f) TR-2. Parameter pairs: (a), (d) $k_{drop,leaf}$ and $k_{leaf,loss}$; (b), (e) $k_{drop,loss}$ and $k_{leaf,loss}$; (c), (f) $k_{drop,store}$ and $k_{drop,leaf}$.
    }
    \label{fig:likelihood-contours-compartmental}
\end{figure}

\begin{figure}[htp]
    \centering
    \begin{subfigure}[b]{0.49\linewidth}
        \centering
        \includegraphics[width=\linewidth]{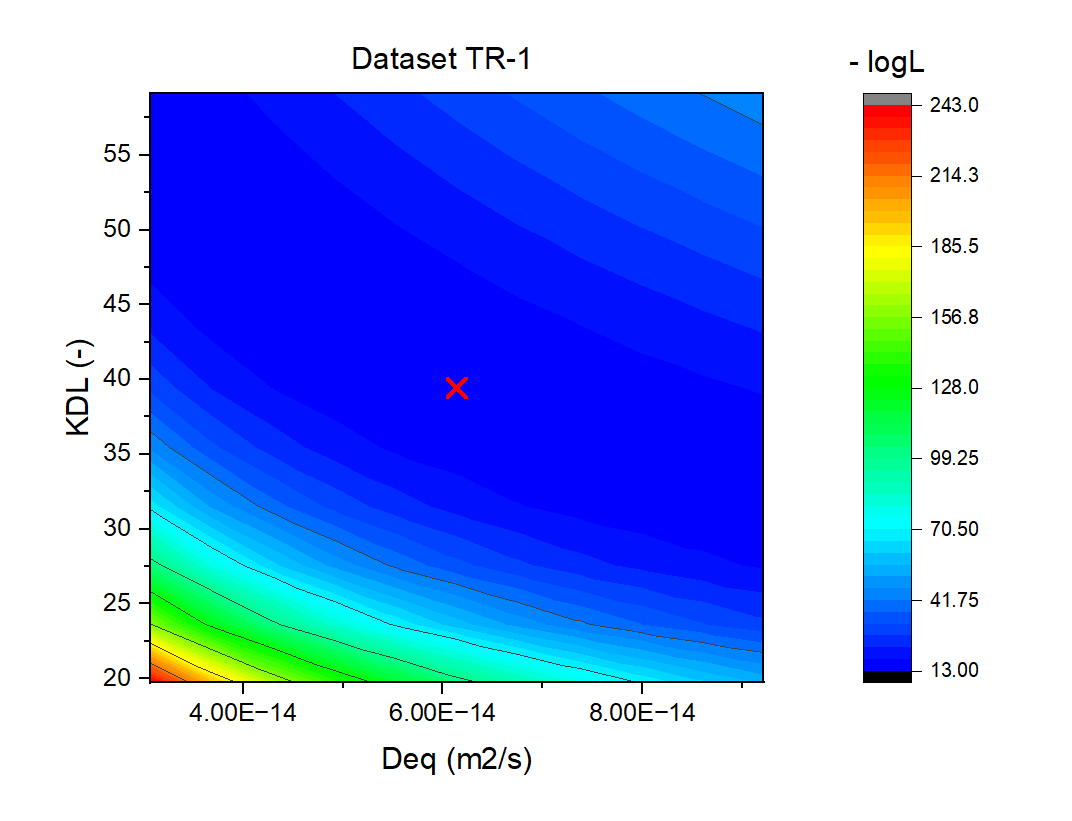}
        \caption{}
        \label{fig:likelihood-contours-diffusion-tr1-d-kdl}
    \end{subfigure}
    \hfill
    \begin{subfigure}[b]{0.49\linewidth}
        \centering
        \includegraphics[width=\linewidth]{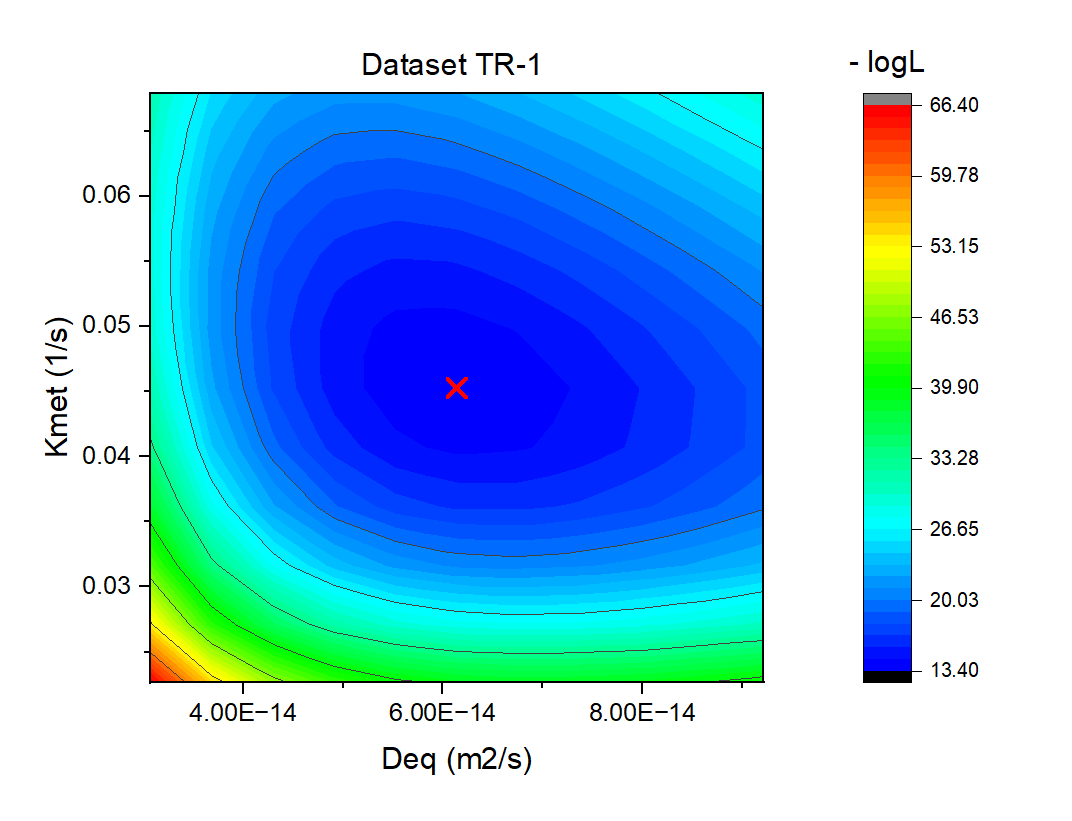}
        \caption{}
        \label{fig:likelihood-contours-diffusion-tr1-d-kmet}
    \end{subfigure}
    \\
    \begin{subfigure}[b]{0.49\linewidth}
        \centering
        \includegraphics[width=\linewidth]{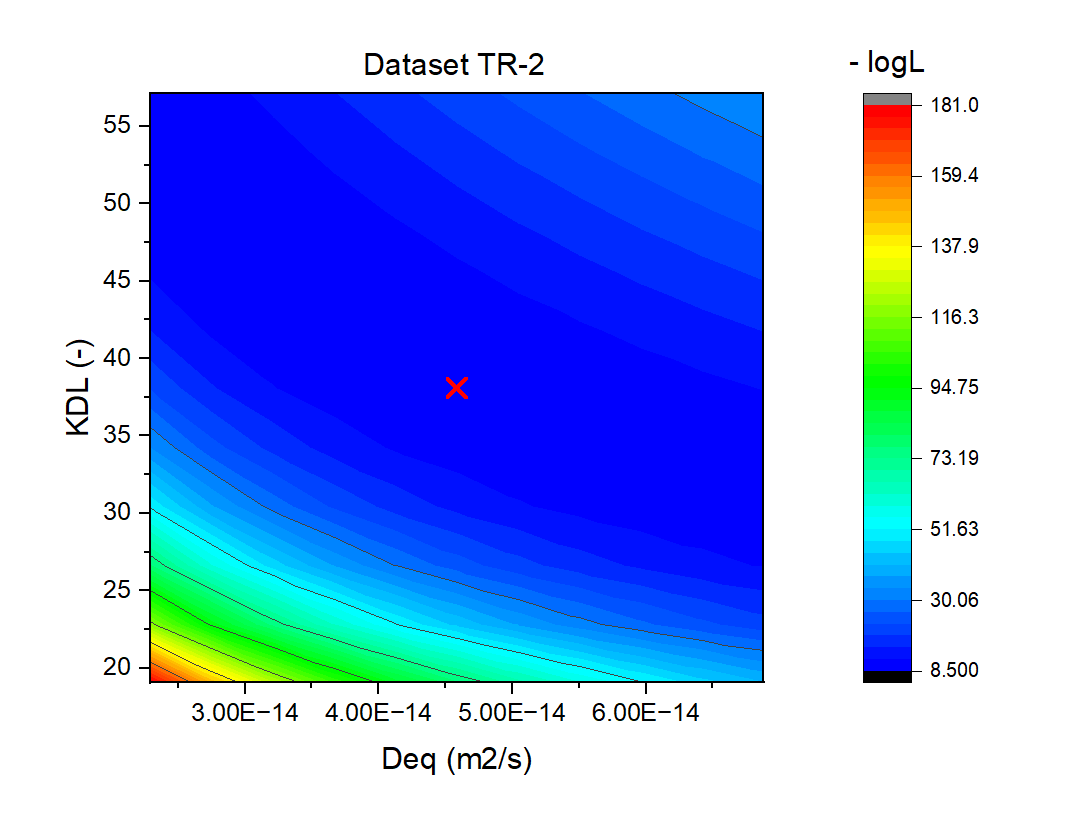}
        \caption{}
        \label{fig:likelihood-contours-diffusion-tr2-d-kdl}
    \end{subfigure}
    \hfill
    \begin{subfigure}[b]{0.49\linewidth}
        \centering
        \includegraphics[width=\linewidth]{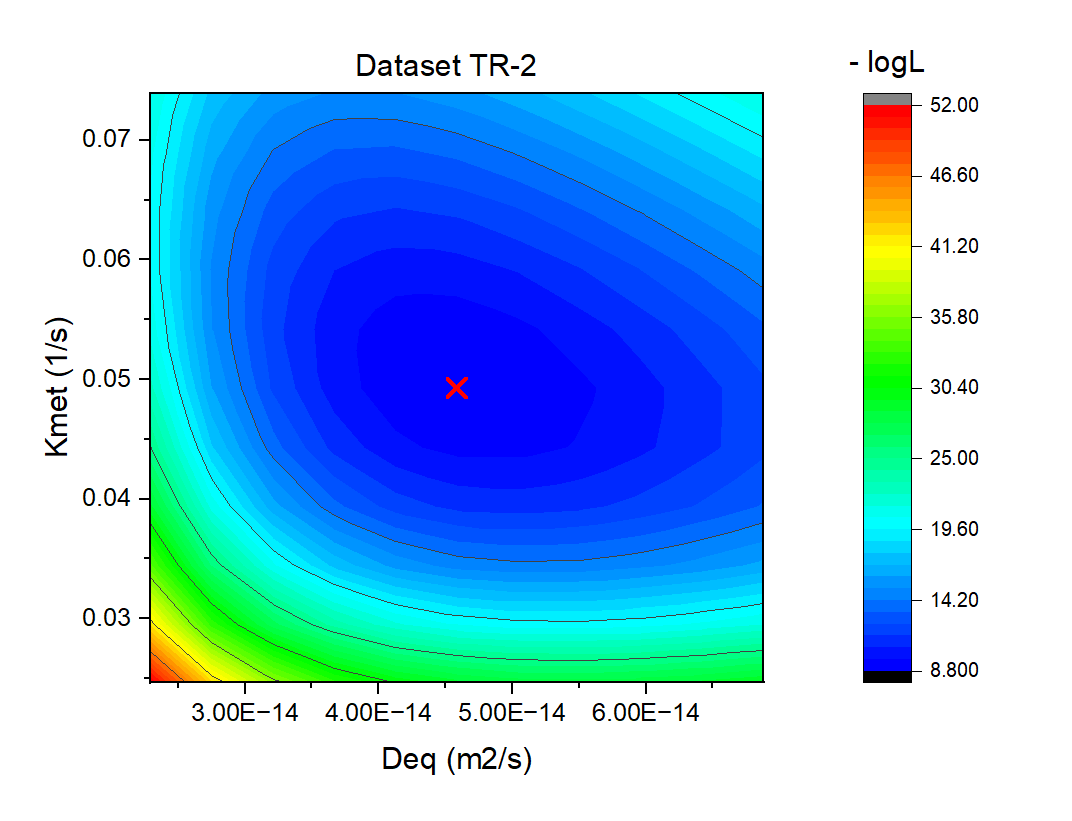}
        \caption{}
        \label{fig:likelihood-contours-diffusion-tr2-d-kmet}
    \end{subfigure}
    \caption{Contour plots showing the likelihood function dependence on pairs of parameters for the diffusion-based model. The red cross is the parameter estimate. Dataset: (a), (b)  TR-1; (c), (d) TR-2. Parameter pairs: (a), (c) $D_{eq}$ and $K_{DL}$; (b), (d) $D_{eq}$ and $K_{met}$.
    }
    \label{fig:likelihood-contours-diffusion}
\end{figure}

This analysis is conducted to observe the correlation profiles for pairs of parameters emerging from the dependence of the likelihood function on the model parameters by means of contour plots. This local study is carried out in the neighbor of the estimate considering a variation of $\pm 50\%$ in the two parameters under assessment, while keeping the other parameters fixed at the estimate.
The study is conducted on both models and with the two datasets TR-1 and TR-2. Figure \ref{fig:likelihood-contours-compartmental} shows the results for the compartmental model, while Figure \ref{fig:likelihood-contours-diffusion} the results for the diffusion model. 

For the diffusion model, plots are reported for the pairs of parameters ($D_{eq}$, $K_{DL}$) and  ($D_{eq}$, $K_{met}$). 
For the compartmental model the following pairs are considered: ($k_{drop,leaf}$, $k_{leaf,loss}$), ($k_{drop,loss}$, $k_{leaf,loss}$), and ($k_{drop,store}$, $k_{drop,leaf}$). 
The pair ($k_{drop,leaf}$, $k_{leaf,loss}$) is selected because these parameters in the compartmental model have a role similar to ($D_{eq}$, $K_{met}$) in the diffusion model, i.e. transport of AI from the droplet to the leaf and consumption of AI within the leaf. The other parameter pairs selected for the compartmental model are those associated with the highest correlation indices in the previous analysis.

The profiles are shown for the negative loglikelihood, therefore the optimum is at the minimum value. The compartmental model plots in Figure~\ref{fig:likelihood-contours-compartmental} show that for the pairs ($k_{drop,leaf}$, $k_{leaf,loss}$) and ($k_{drop,loss}$, $k_{leaf,loss}$) the optimum is well identified locally, while the region of maximum likelihood is broader when reducing the parameter space to ($k_{drop,store}$, $k_{drop,leaf}$) in Figure~\ref{fig:likelihood-contours-compartmental-tr1-c} and \ref{fig:likelihood-contours-compartmental-tr2-c}. For these two parameters, the 0.91 correlation obtained before seems to have a linear dependence around the estimate.

The plots for the diffusion model show that the maximum likelihood is well identified when considering the pair ($D_{eq}$, $K_{met}$), Figure~\ref{fig:likelihood-contours-diffusion-tr1-d-kmet} and \ref{fig:likelihood-contours-diffusion-tr2-d-kmet}. An interesting pattern in the likelihood profiles emerges in Figure~\ref{fig:likelihood-contours-diffusion-tr1-d-kdl} and \ref{fig:likelihood-contours-diffusion-tr2-d-kdl}, when the parameters under assessment are ($D_{eq}$, $K_{DL}$). These parameters are totally anti-correlated based on the results of the analysis presented before, and from this study the correlation between $D_{eq}$ and $K_{DL}$ can be observed having a nonlinear trend. 

\subsection{Data augmentation}  \label{sec:results-data}

Following the procedure presented in the methodological framework, the last analysis presented in this paper is the data augmentation study. The two different data augmentation strategies described in the methodology Section~\ref{sec:method-data-augmentation} are presented:
\begin{itemize}
    \item Strategy 1 - single additional sample.\\
    Study conducted to assess the expected improvement in parameter estimation identifiability depending on the experimental design $\bm{\varphi}$, i.e. sampling time in this case.
    \item Strategy 2 - multiple additional samples.\\
    Analysis to verify how many additional data are required to solve parameter identifiability issues. 
\end{itemize}

Results from the two different augmentation strategies are reported in the following subsections.

\subsubsection{First augmented data study}
\label{sec:res-data-aug-first}

The first analysis considers the effect of adding one additional experimental data to the 8 data points in the original dataset to assess the expected improvement in the statistics of parameter estimation, data fitting and identifiability of model parameters. In this study, the impact of the different location in time for the additional sample is evaluated. The noise factor added to the simulated measurement is generated with a heteroscedastic model calibrated on the noise observed in the original dataset, to replicate the data variability of the real experiments. This study is conducted on both available datasets TR-1 and TR-2, and the heteroscedastic model is calibrated independently for each case.

\begin{figure}[htp]
    \centering
    \begin{subfigure}[b]{0.49\linewidth}
        \centering
        \includegraphics[width=\linewidth]{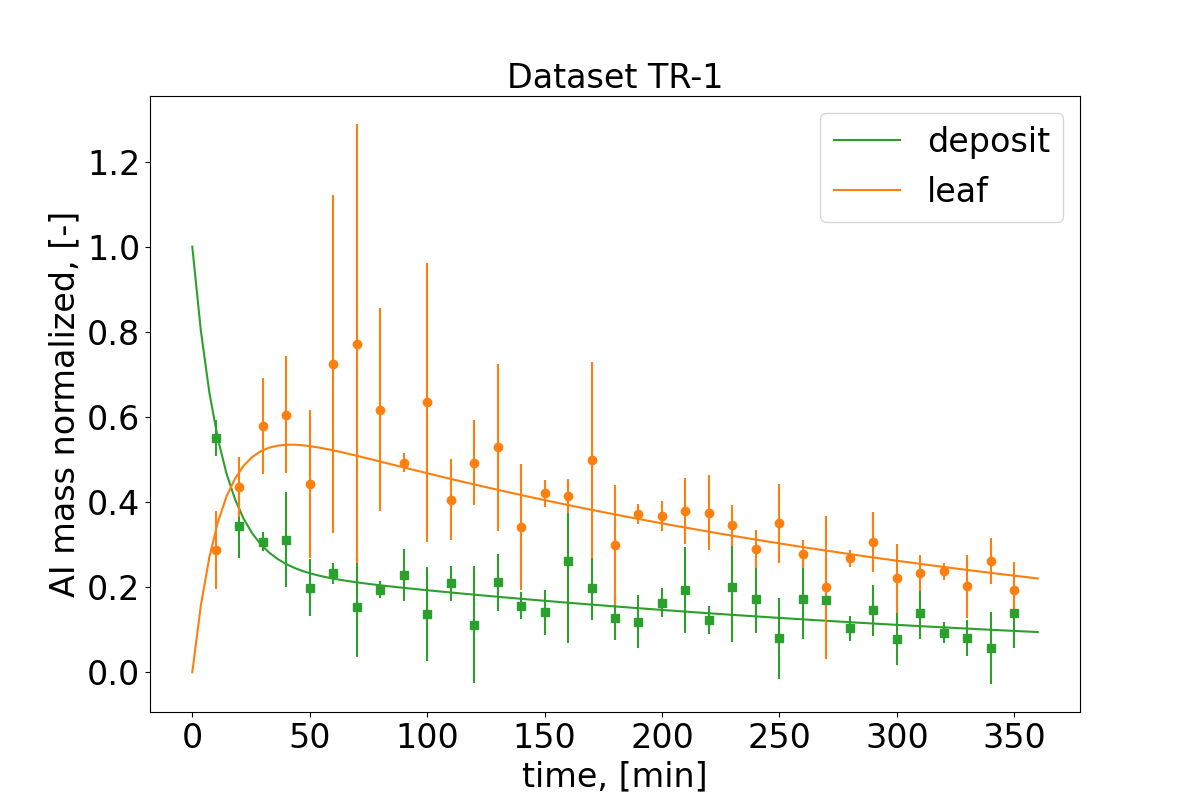}
        \caption{}
    \end{subfigure}
    \hfill
    \begin{subfigure}[b]{0.49\linewidth}
        \centering
        \includegraphics[width=\linewidth]{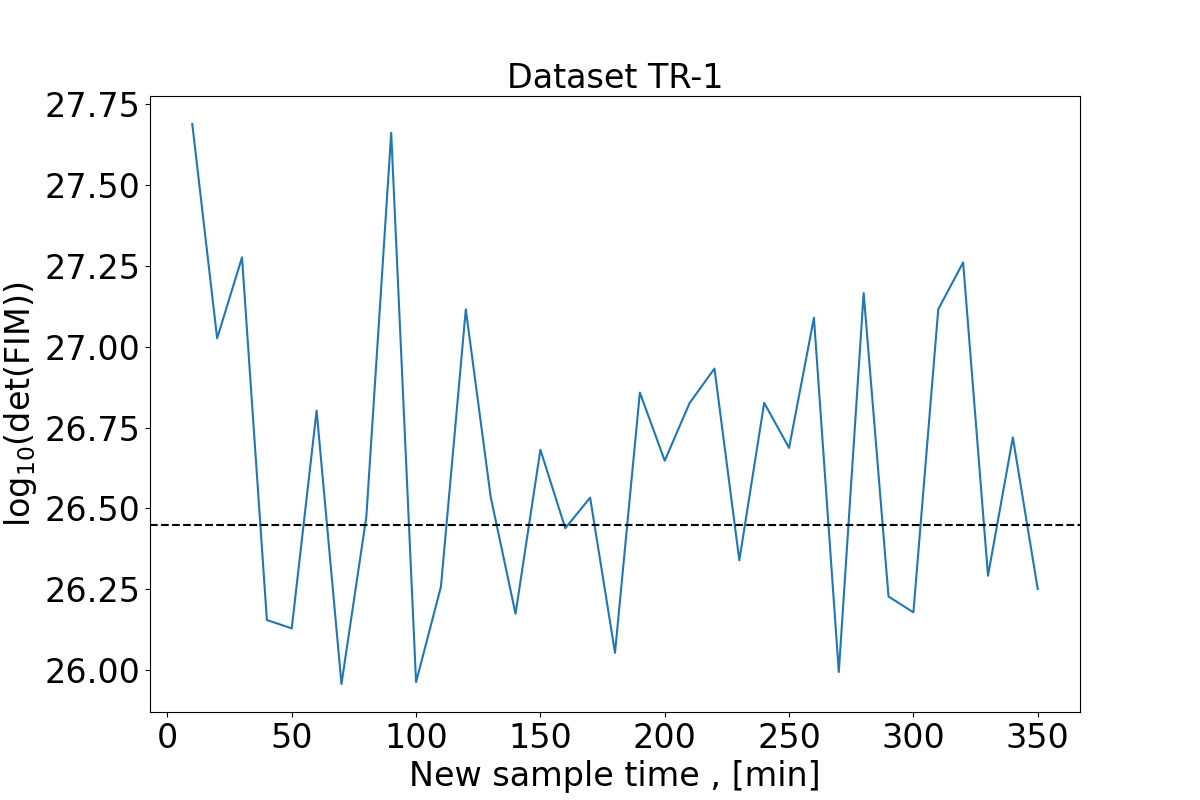}
        \caption{}
    \end{subfigure}
    \\
    \begin{subfigure}[b]{0.49\linewidth}
        \centering
        \includegraphics[width=\linewidth]{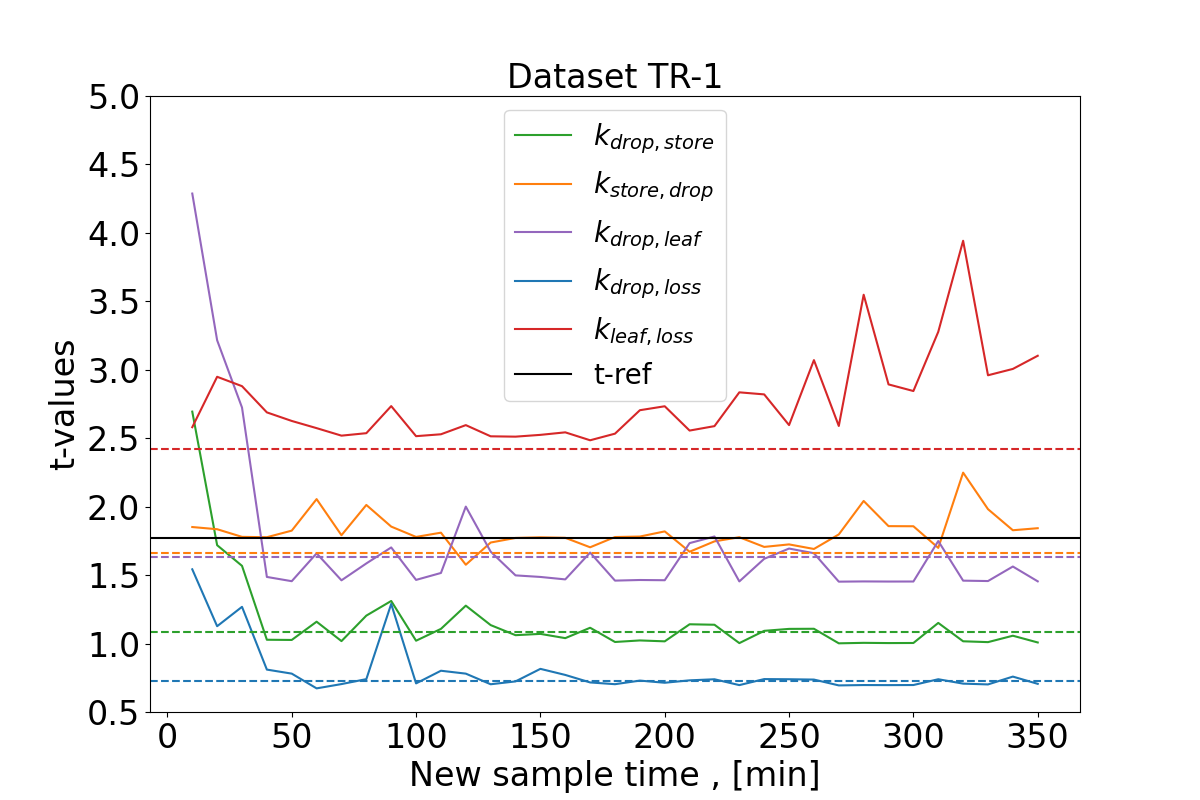}
        \caption{}
        \label{fig:data-augm-comp-TR1-tval}
    \end{subfigure}
    \hfill
    \begin{subfigure}[b]{0.49\linewidth}
        \centering
        \includegraphics[width=\linewidth]{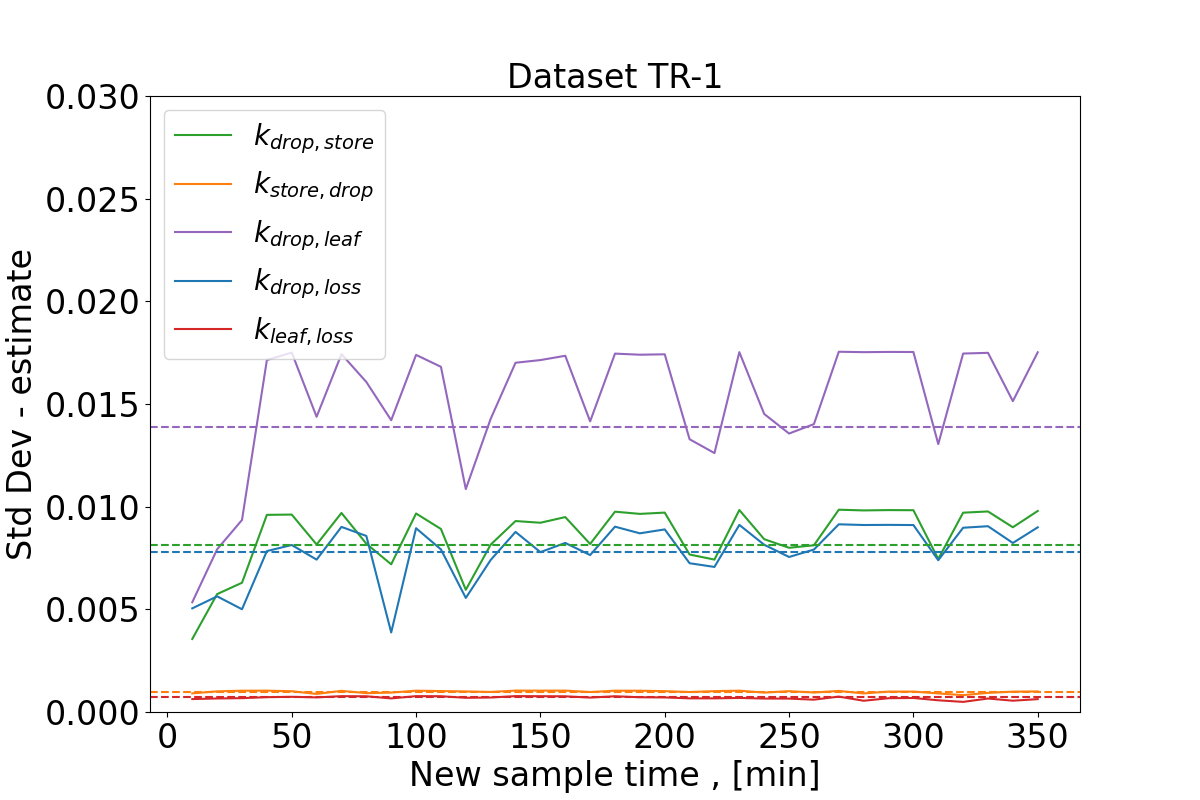}
        \caption{}
    \end{subfigure}    \\
    \begin{subfigure}[b]{0.49\linewidth}
        \centering
        \includegraphics[width=\linewidth]{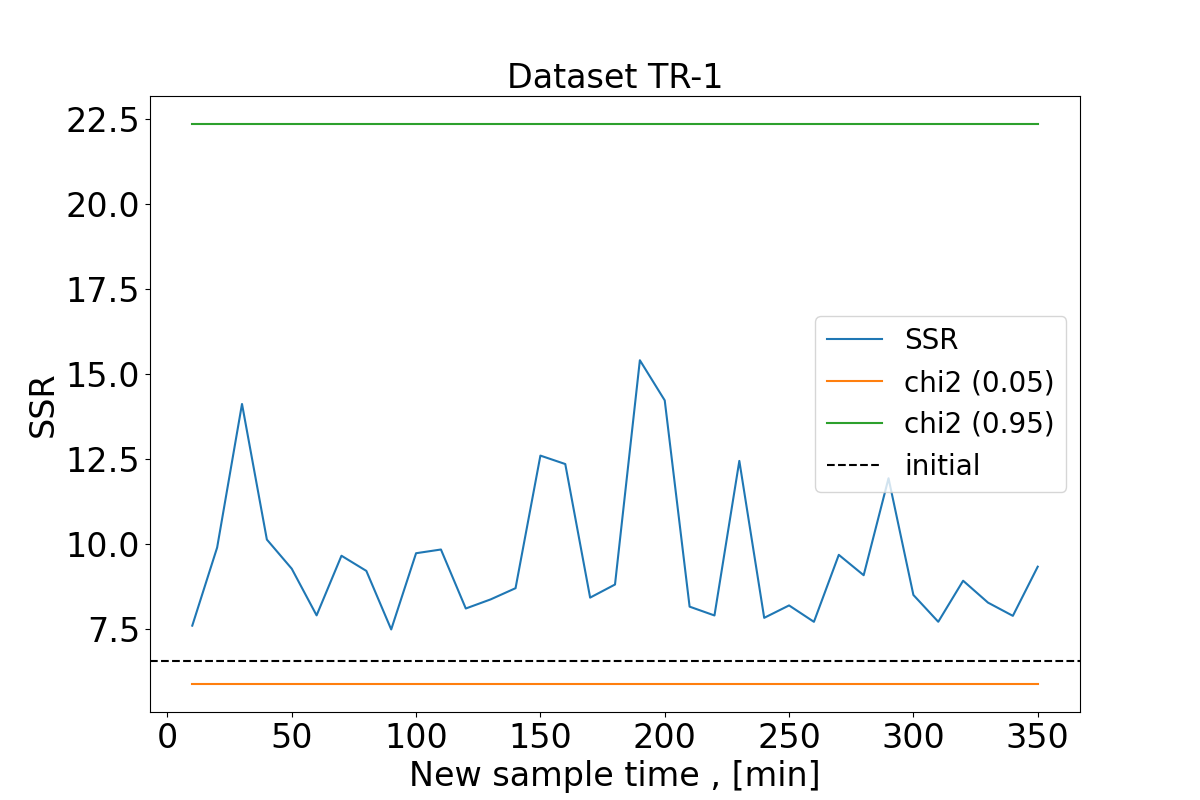}
        \caption{}
        \label{fig:data-augm-comp-TR1-SSR}
    \end{subfigure}

    \caption{Data augmentation results for the compartmental model, dataset TR-1. (a) Simulated data with the experimental error, (b) det(FIM) - log$_{10}$ scale calculated as in Eq.~\ref{eq:fim}, (c) t-values, (d) standard deviation in the estimates, (e) SSR after data fitting. Dashed lines show the values obtained with the original dataset. 
    }
    \label{fig:data-augm-comp-TR1}
\end{figure}

\begin{figure}[htp]
    \centering
    \begin{subfigure}[b]{0.49\linewidth}
        \centering
        \includegraphics[width=\linewidth]{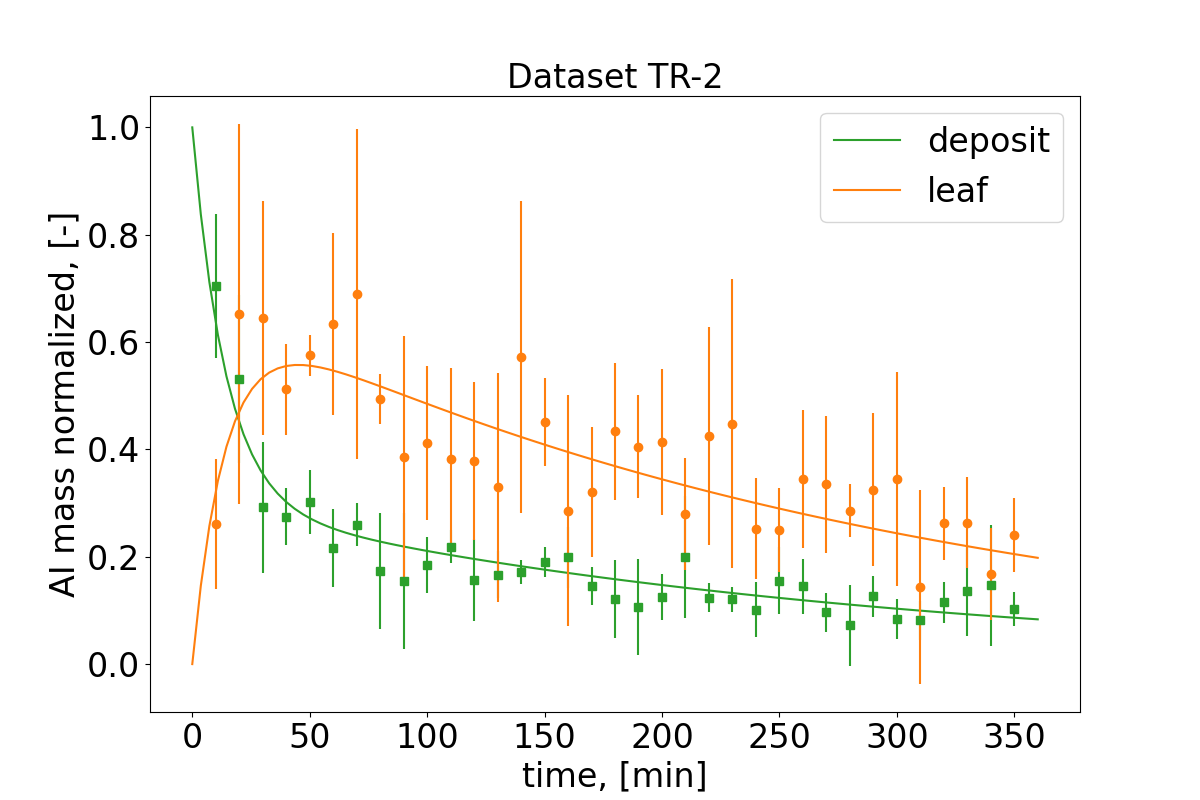}
        \caption{}
    \end{subfigure}
    \hfill
    \begin{subfigure}[b]{0.49\linewidth}
        \centering
        \includegraphics[width=\linewidth]{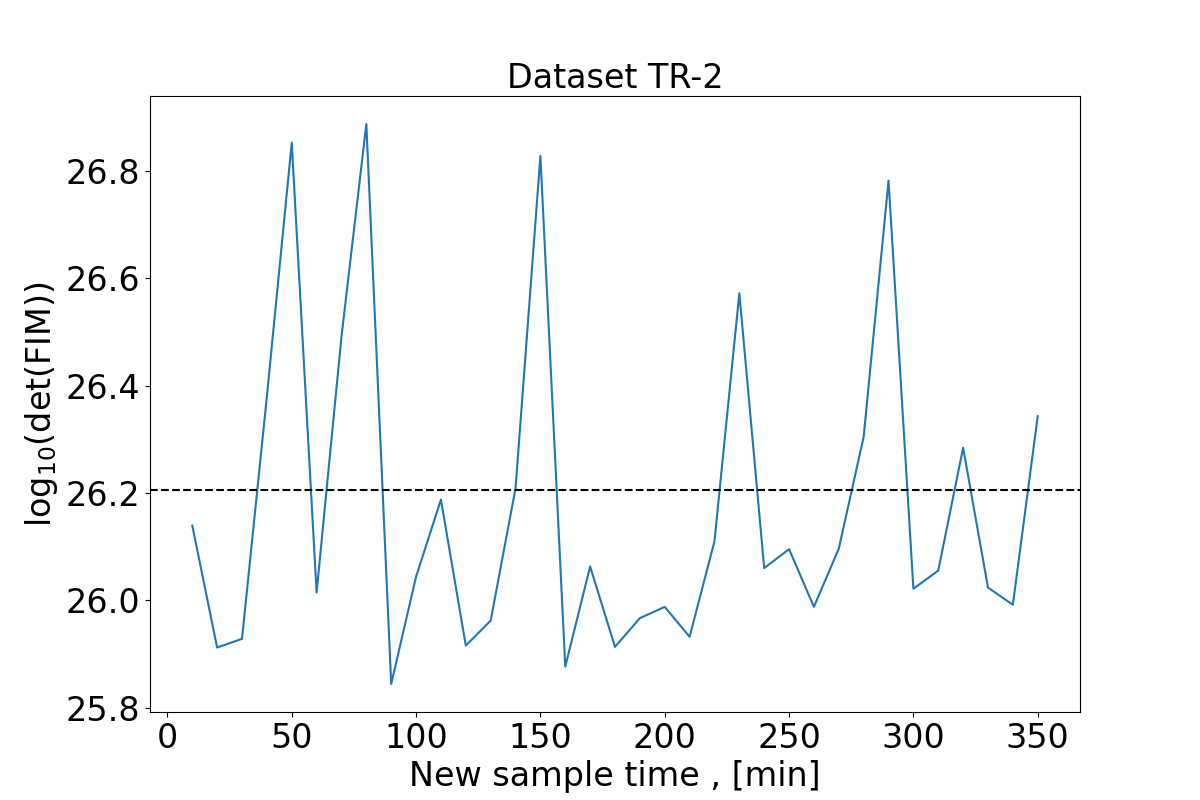}
        \caption{}
    \end{subfigure}
    \\
    \begin{subfigure}[b]{0.49\linewidth}
        \centering
        \includegraphics[width=\linewidth]{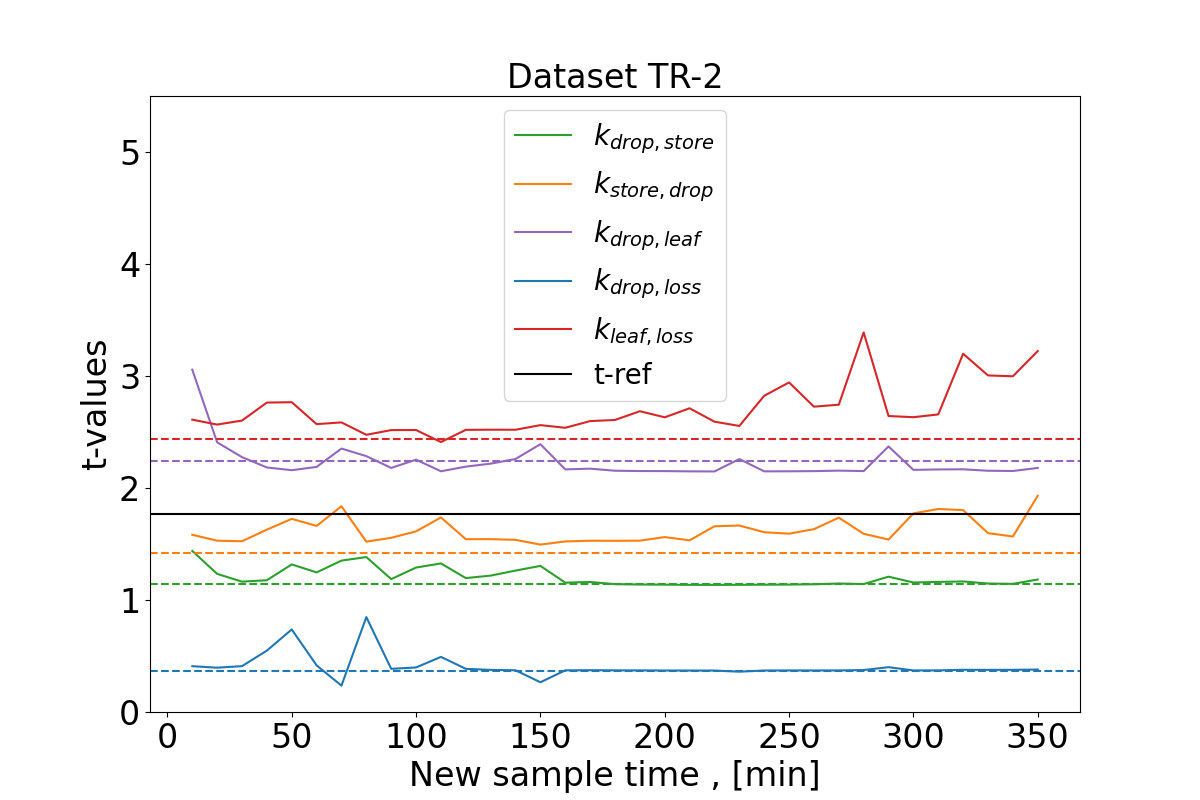}
        \caption{}
        \label{fig:data-augm-comp-TR2-tval}
    \end{subfigure}
    \hfill
    \begin{subfigure}[b]{0.49\linewidth}
        \centering
        \includegraphics[width=\linewidth]{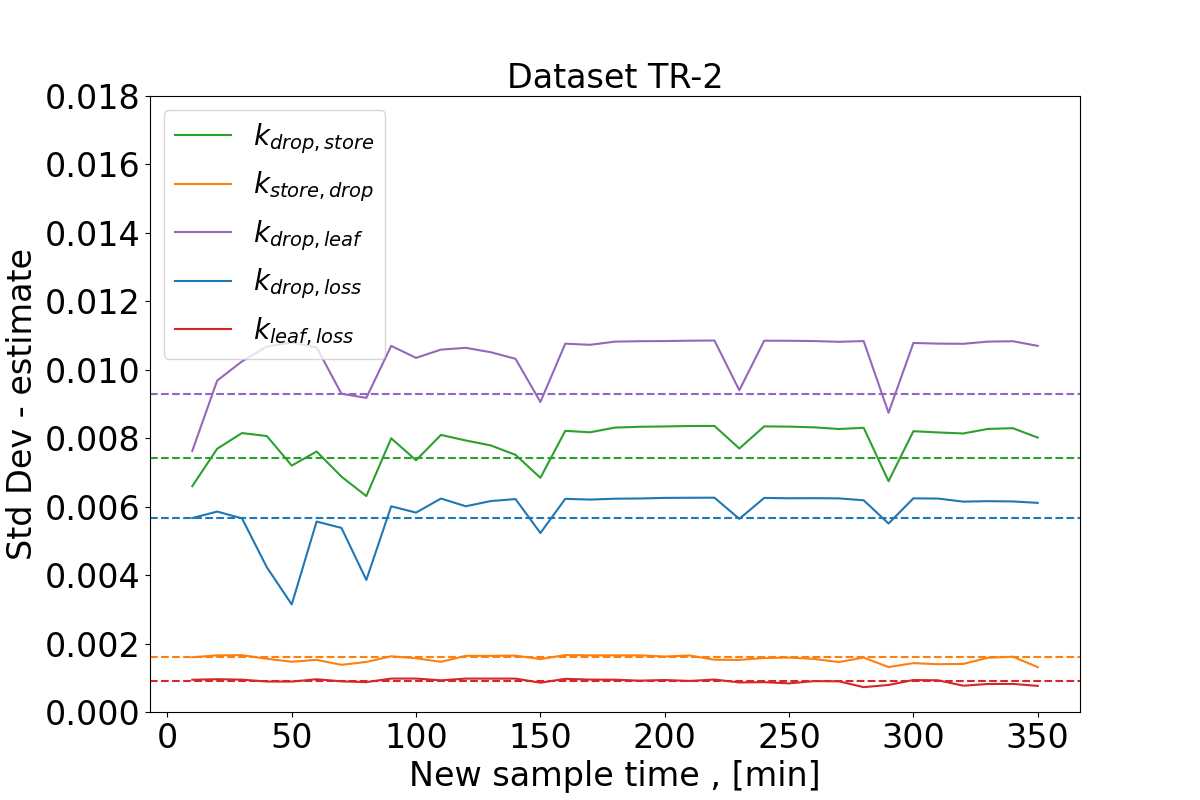}
        \caption{}
    \end{subfigure}    \\
    \begin{subfigure}[b]{0.49\linewidth}
        \centering
        \includegraphics[width=\linewidth]{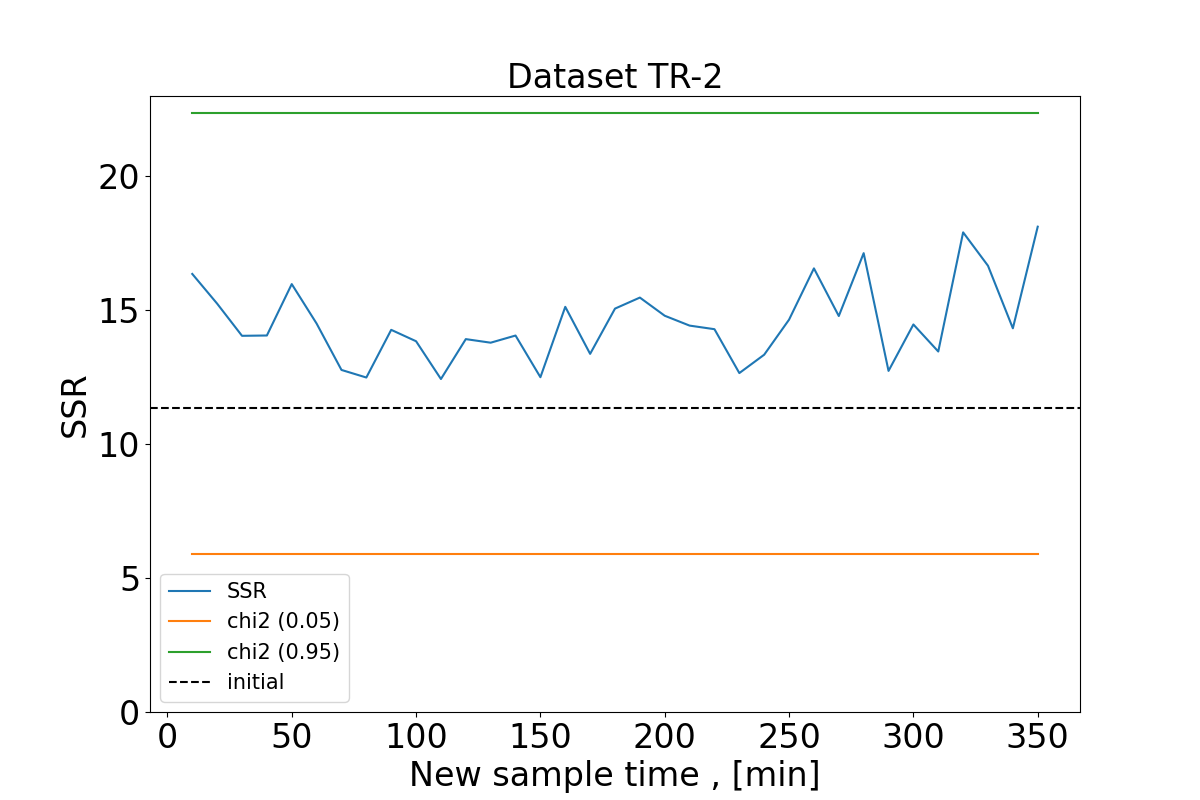}
        \caption{}
        \label{fig:data-augm-comp-TR2-SSR}
    \end{subfigure}
    \caption{Data augmentation results for the compartmental model, dataset TR-2. (a) Simulated data with the experimental error, (b) det(FIM) - log$_{10}$ scale calculated as in Eq.~\ref{eq:fim}, (c) t-values, (d) standard deviation in the estimates, (e) SSR after data fitting. Dashed lines show the values obtained with the original dataset.
    }
    \label{fig:data-augm-comp-TR2}
\end{figure}

\begin{figure}[htp]
    \centering
    \begin{subfigure}[b]{0.49\linewidth}
        \centering
        \includegraphics[width=\linewidth]{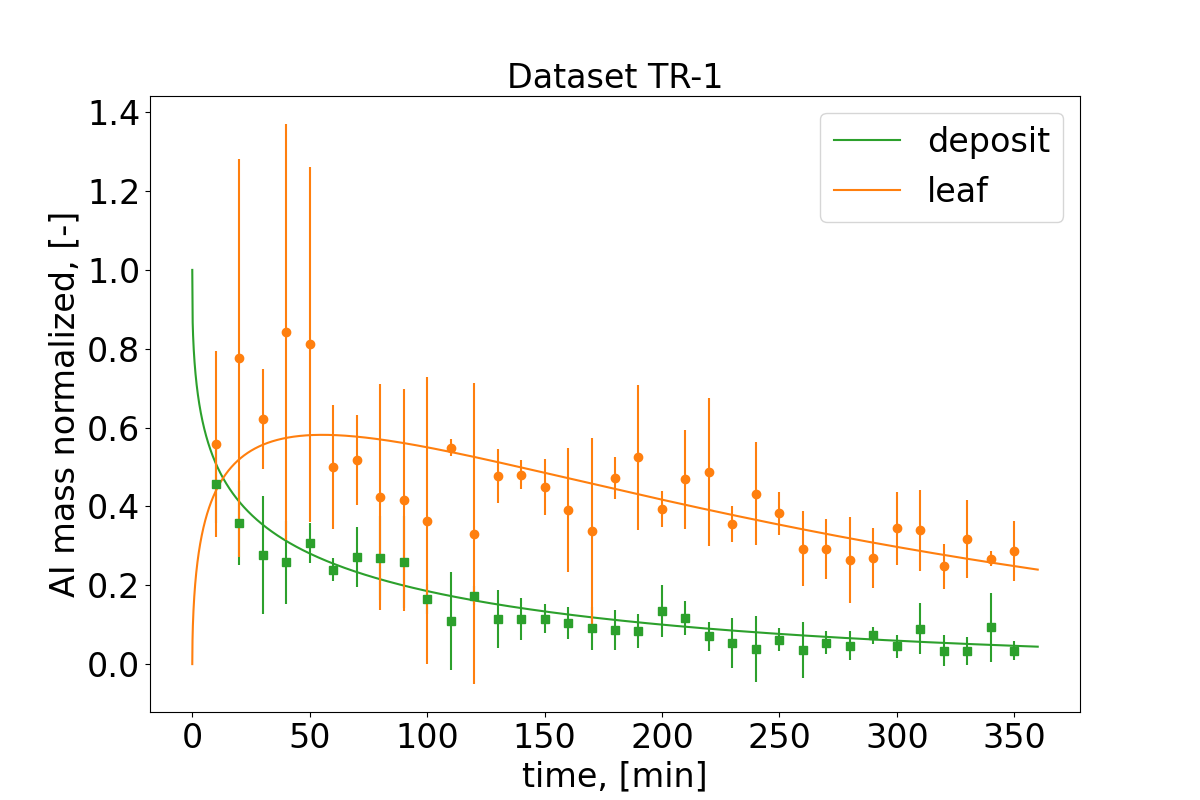}
        \caption{}
    \end{subfigure}
    \hfill
    \begin{subfigure}[b]{0.49\linewidth}
        \centering
        \includegraphics[width=\linewidth]{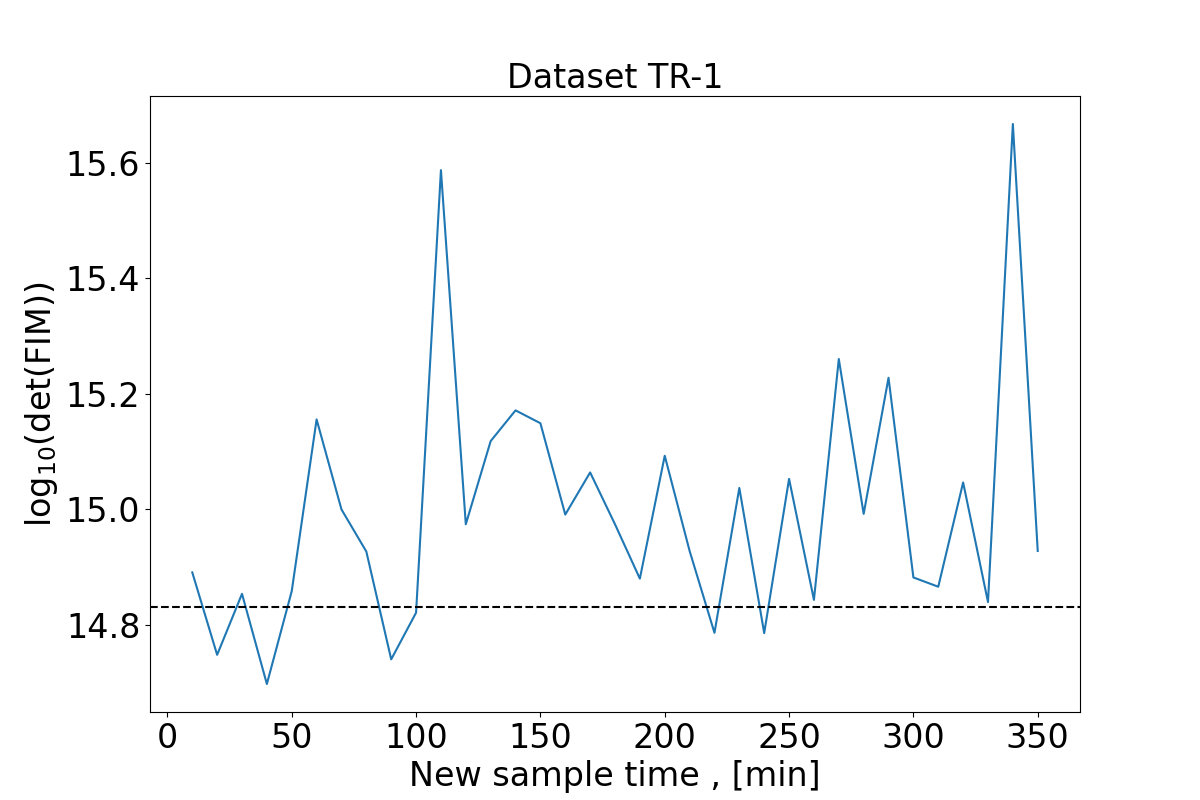}
        \caption{}
    \end{subfigure}
    \\
    \begin{subfigure}[b]{0.49\linewidth}
        \centering
        \includegraphics[width=\linewidth]{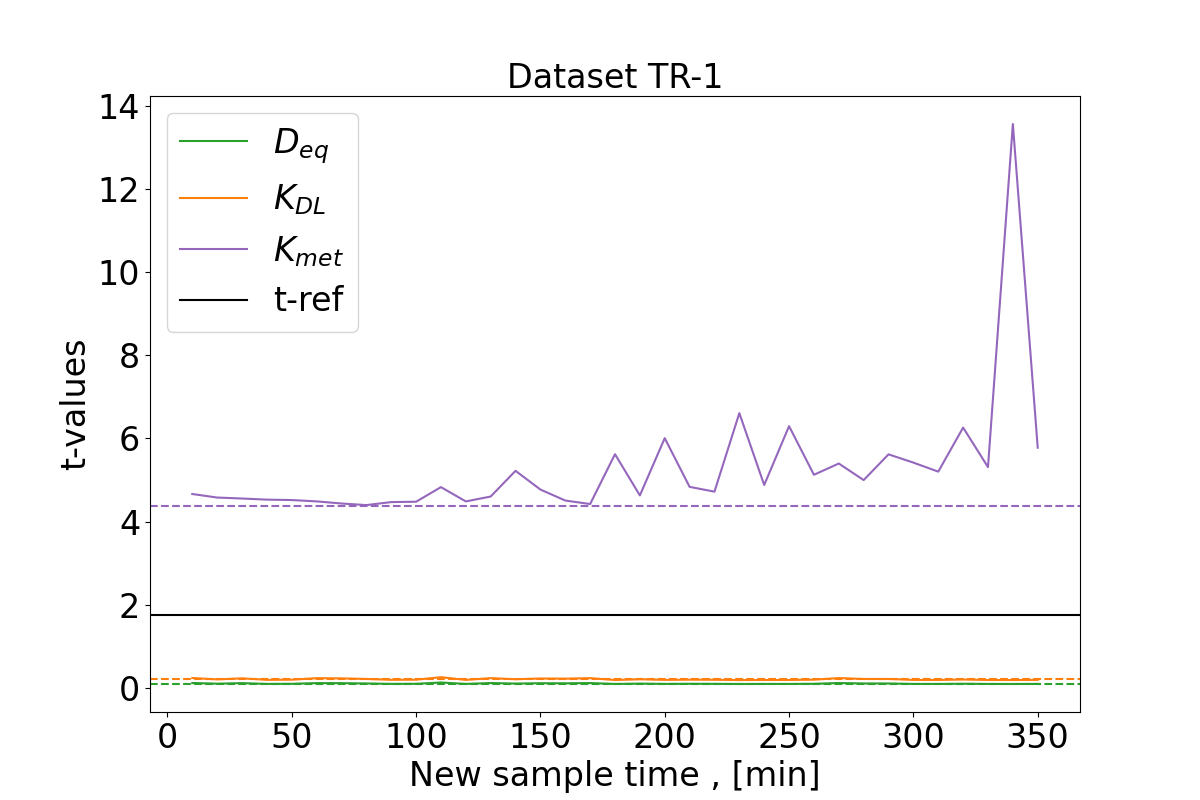}
        \caption{}
        \label{fig:data-augm-diff-TR1-tval}
    \end{subfigure}
    \hfill
    \begin{subfigure}[b]{0.49\linewidth}
        \centering
        \includegraphics[width=\linewidth]{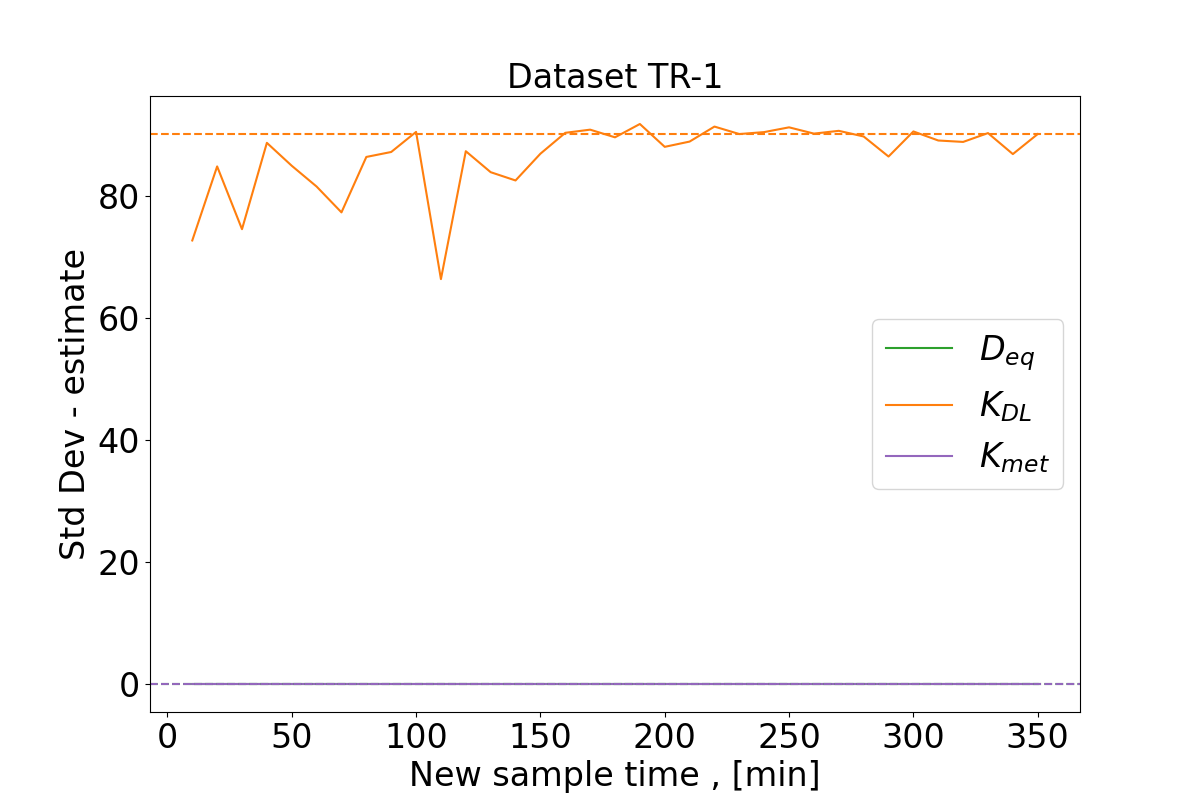}
        \caption{}
    \end{subfigure}    \\
    \begin{subfigure}[b]{0.49\linewidth}
        \centering
        \includegraphics[width=\linewidth]{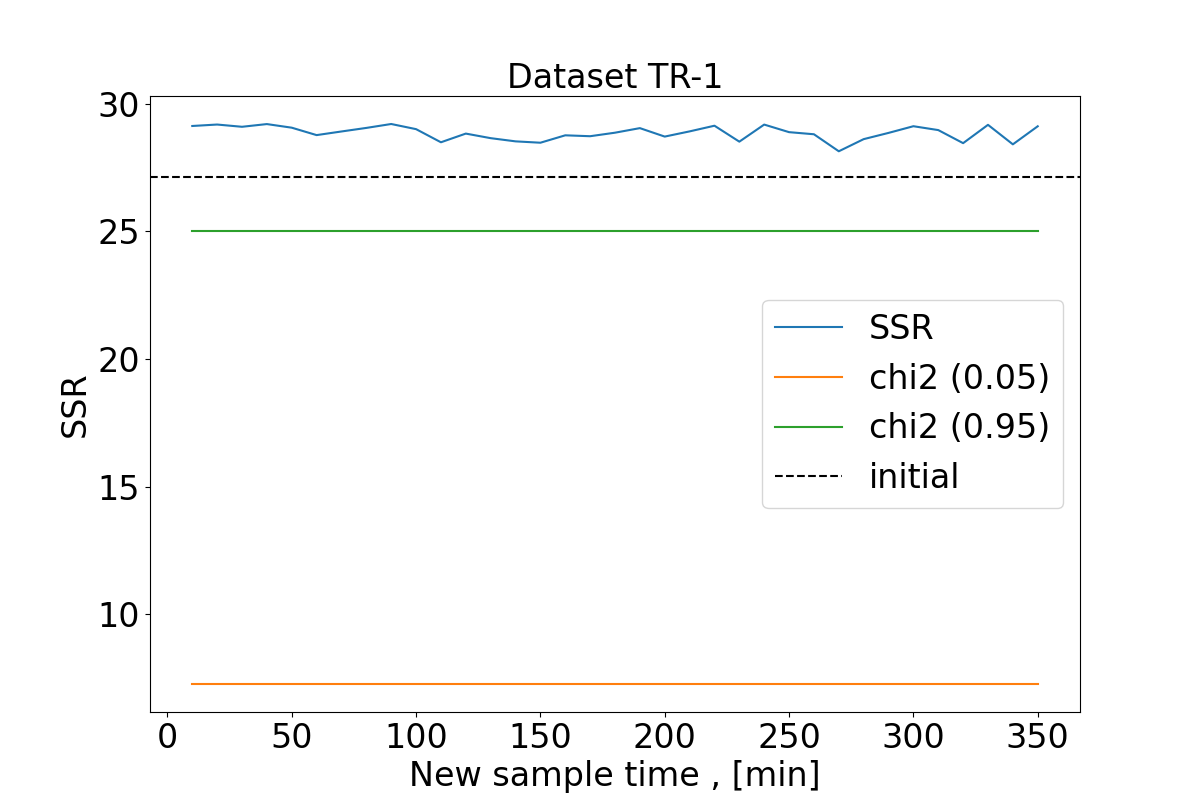}
        \caption{}
        \label{fig:data-augm-diff-TR1-SSR}
    \end{subfigure}
    \caption{Data augmentation results for the diffusion-based model, dataset TR-1. (a) Simulated data with the experimental error, (b) det(FIM) - log$_{10}$ scale calculated as in Eq.~\ref{eq:fim}, (c) t-values, (d) standard deviation in the estimates, (e) SSR after data fitting. Dashed lines show the values obtained with the original dataset. 
    }
    \label{fig:data-augm-diff-TR1}
\end{figure}

\begin{figure}[htp]
    \centering
    \begin{subfigure}[b]{0.49\linewidth}
        \centering
        \includegraphics[width=\linewidth]{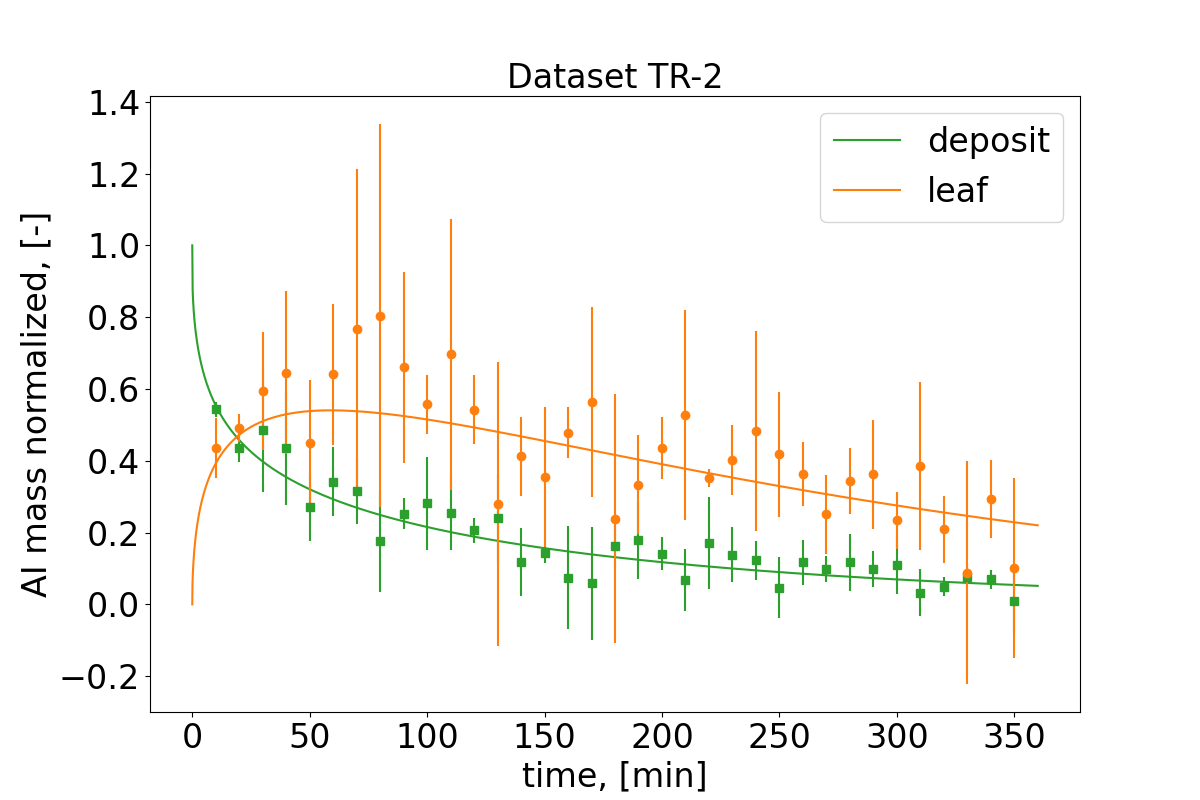}
        \caption{}
    \end{subfigure}
    \hfill
    \begin{subfigure}[b]{0.49\linewidth}
        \centering
        \includegraphics[width=\linewidth]{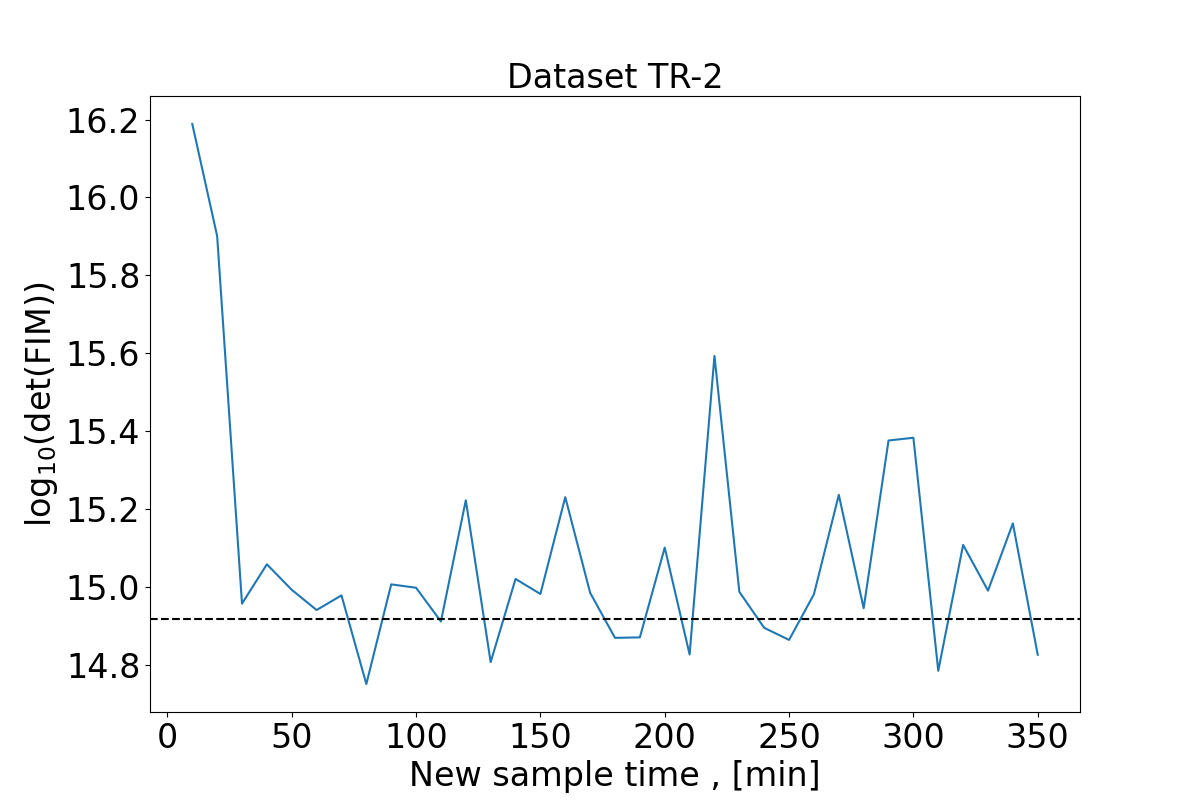}
        \caption{}
    \end{subfigure}
    \\
    \begin{subfigure}[b]{0.49\linewidth}
        \centering
        \includegraphics[width=\linewidth]{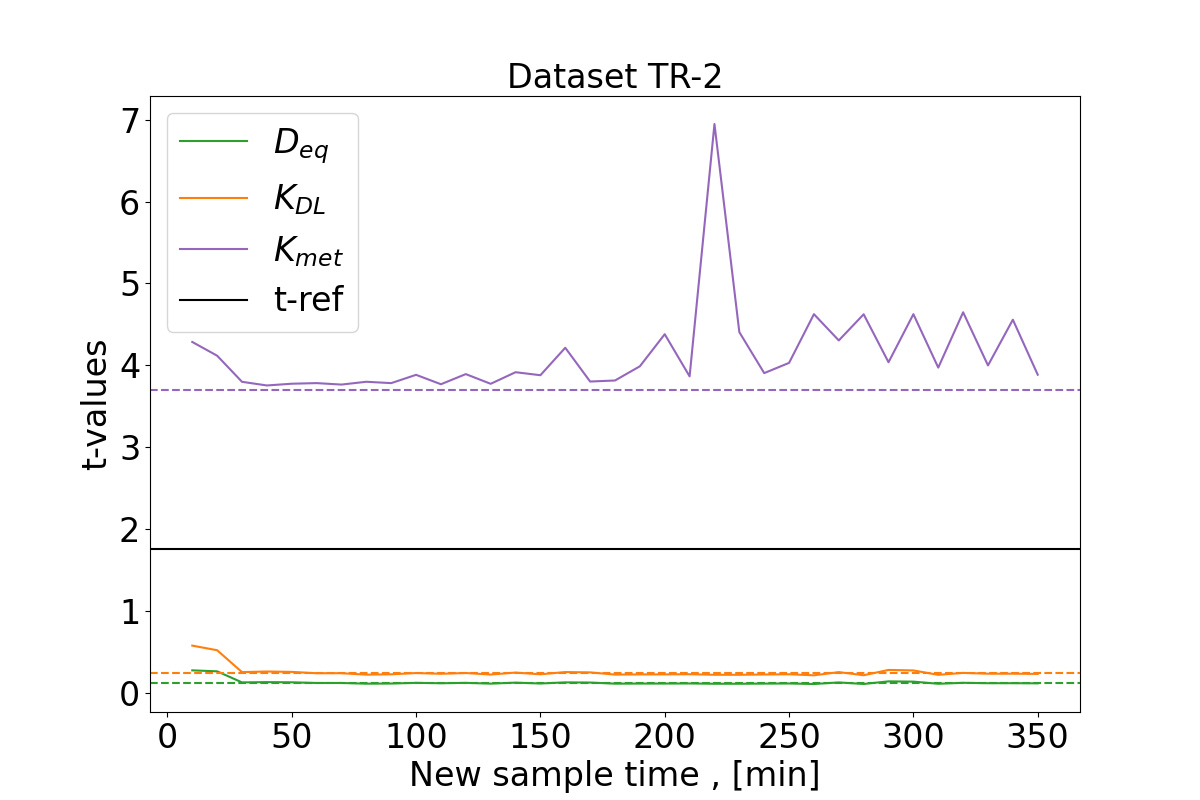}
        \caption{}
        \label{fig:data-augm-diff-TR2-tval}
    \end{subfigure}
    \hfill
    \begin{subfigure}[b]{0.49\linewidth}
        \centering
        \includegraphics[width=\linewidth]{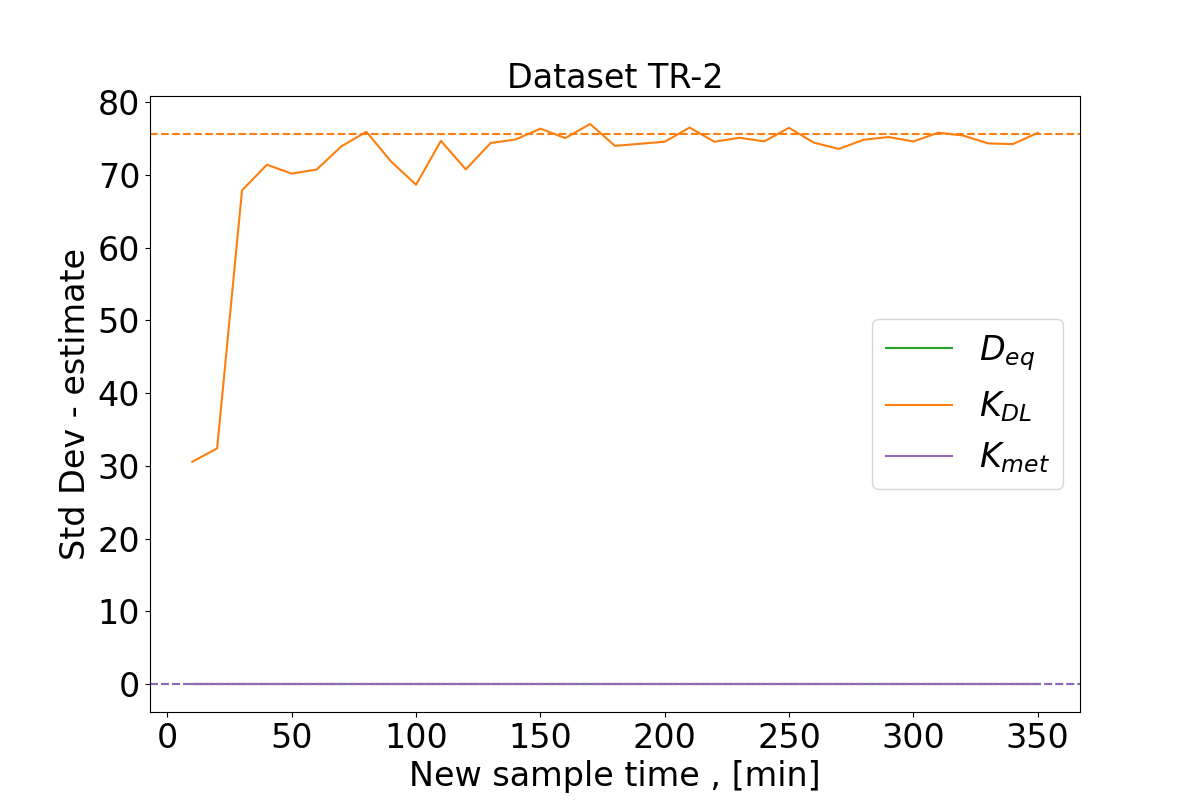}
        \caption{}
    \end{subfigure}    \\
    \begin{subfigure}[b]{0.49\linewidth}
        \centering
        \includegraphics[width=\linewidth]{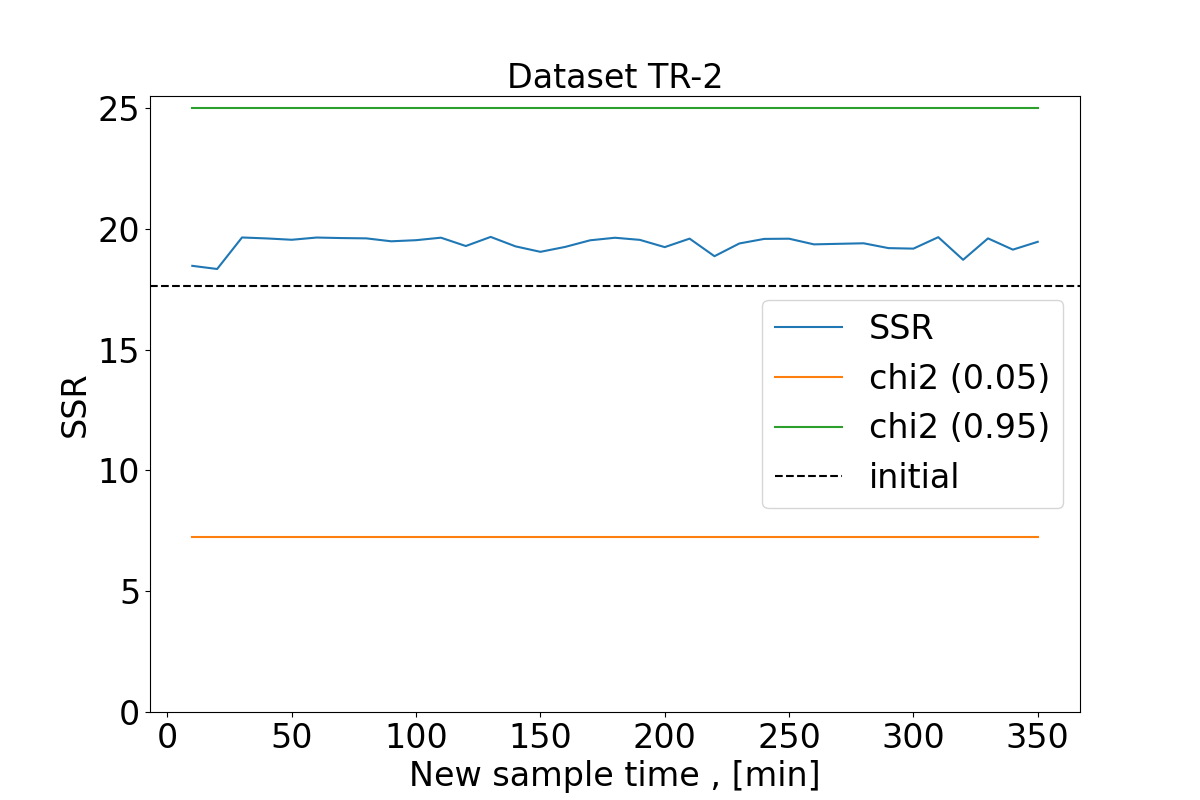}
        \caption{}
        \label{fig:data-augm-diff-TR2-SSR}
    \end{subfigure}
    \caption{Data augmentation results for the diffusion-based model, dataset TR-2. (a) Simulated data with the experimental error, (b) det(FIM) - log$_{10}$ scale calculated as in Eq.~\ref{eq:fim}, (c) t-values, (d) standard deviation in the estimates, (e) SSR after data fitting. Dashed lines show the values obtained with the original dataset.  
    }
    \label{fig:data-augm-diff-TR2}
\end{figure}

Results are shown in Figure \ref{fig:data-augm-comp-TR1} for the compartmental model and TR-1, Figure \ref{fig:data-augm-comp-TR2} for the compartmental model and TR-2, Figure \ref{fig:data-augm-diff-TR1} for the diffusion model and TR-1, Figure \ref{fig:data-augm-diff-TR2} for the diffusion model and TR-2.
For each combination of model and dataset the following plots are reported: (a) the whole set of additional experimental data simulated, and the profiles with respect to the sampling time of the additional data point of (b) $\log_{10}(\det(FIM))$, (c) the t-value statistics for the parameters, (d) the standard deviation of the estimates, (e) the SSR with the $\chi^2$ reference values. Dashed lines in the plots indicate the initial values obtained with the original dataset before adding the new simulated data point.

The profiles of the determinant of the FIM depict oscillations around the initial value for the compartmental model. This can be due to the high variability in experimental noise added to the new data, which is included in the calculation of FIM in the term $\bm{\Sigma}_y$ as per Equation~\eqref{eq:fim}, and since the dataset dimensionality is small, just a single additional data point can have a significant impact. It is noted that for the diffusion-based model the det(FIM) trend increases from the original value with the new experimental data regardless of the sampling time, still with oscillations. 

The SSR increases for all the models/treatments with the new data point, but it remains within the statistically significant region ($\chi^2_{0.05}-\chi^2_{0.95}$) in Figure~\ref{fig:data-augm-comp-TR1-SSR}, \ref{fig:data-augm-comp-TR2-SSR} and \ref{fig:data-augm-diff-TR2-SSR}. For the diffusion model, Fig.~\ref{fig:data-augm-diff-TR1-SSR}, the quality of fitting on dataset TR-1 is still poor due to under-fitting (SSR > $\chi^2_{0.95}$) even with the new data point, as it was with the original dataset. 

The quality of the estimates obtained with the new data is shown with the plots of t-test statistics and of the standard deviation of the estimates. These two profiles are related because the lower the parametric uncertainty, the higher the t-value, for a given estimate of the parameter, as per Eq.~\eqref{eq:tvalue}.
For the compartmental model, adding a data point collected in the first 30 minutes can significantly improve the quality of the estimates, as shown in Figure~\ref{fig:data-augm-comp-TR1-tval}, especially for the parameters $k_{drop,leaf}$, $k_{drop,store}$ and $k_{drop,loss}$, which were characterized by the lowest t-values. The early location in time of the most informative sample for these parameters is in agreement with the sensitivity profiles presented in Figure~\ref{fig:sensitivity-comp}.

The results obtained with the diffusion-based model are also in agreement with the sensitivity study shown in Figure~\ref{fig:sensitivity-diff}. Clear directions for experimentation are drawn from these results: an increase in the quality of the estimate for $K_{met}$ is observed with samples collected at times longer than 3 hours, while the confidence in the estimate for the parameters that have identifiability issues ($K_{DL}$ and $D_{eq}$) is slightly improved by adding a new sample collected in the short times, i.e. within 2 hours from the droplet deposition on the leaves, with a peak in the first 30 minutes. However, for these two parameters the improvement obtained with a single data point is minimal and the t-values are still far from the threshold reference value. 

\subsubsection{Second augmented data study} \label{sec:res-data-aug-second}

The questions arising from the first data augmentation study in Section~\ref{sec:res-data-aug-first} are then: if more budget is potentially available to collect new samples from the same type of experiments, would it be possible to obtain a good estimation for these two parameters? And if so, how many experiments would be required?

To answer these questions, the second data augmentation study described in the methodology (Section~\ref{sec:method-data-augmentation}) is performed. This analysis is conducted only for the diffusion-based model since it is the only model with practical identifiability problems. It is assumed to collect the new data in the time interval between 10 and 120 minutes, and to have a budget that goes up from 0 to 150 new data points equally spaced in this time frame.

\begin{figure}[htp]
    \centering
    \begin{subfigure}[b]{0.49\linewidth}
        \centering
        \includegraphics[width=\linewidth]{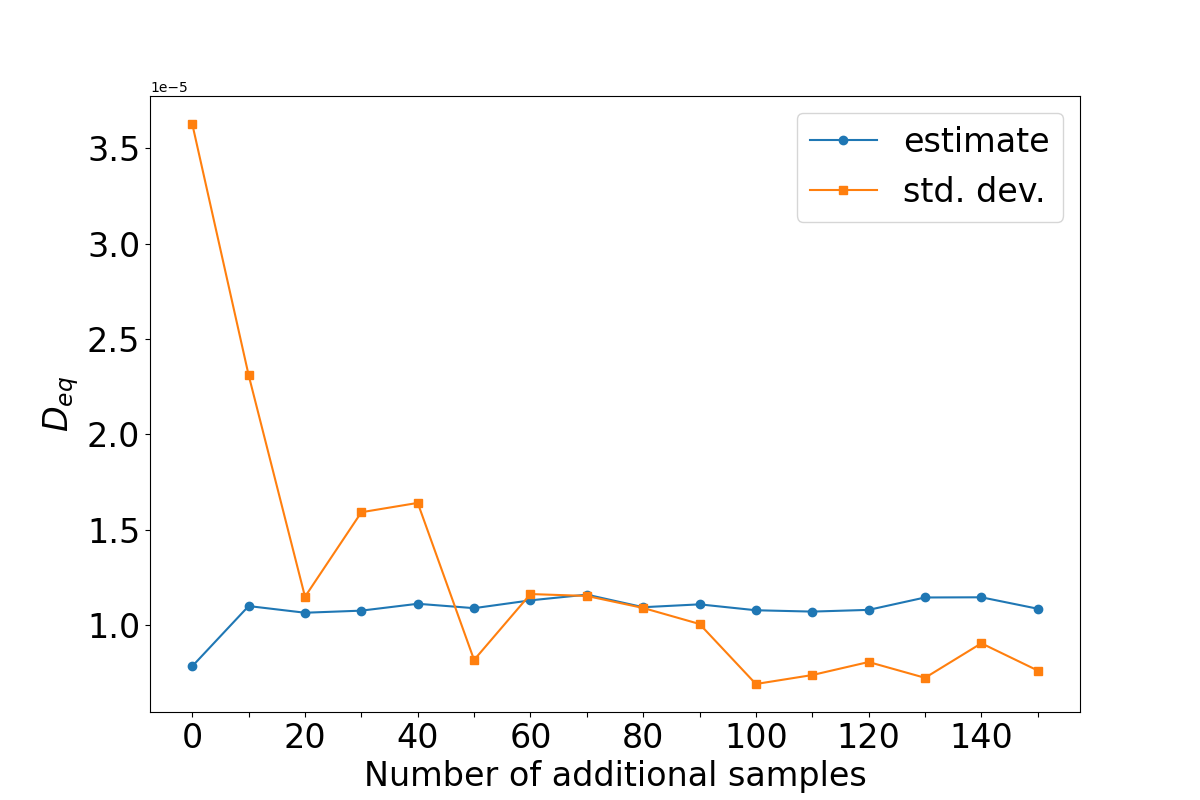}
        \caption{}
    \end{subfigure}
    \hfill
    \begin{subfigure}[b]{0.49\linewidth}
        \centering
        \includegraphics[width=\linewidth]{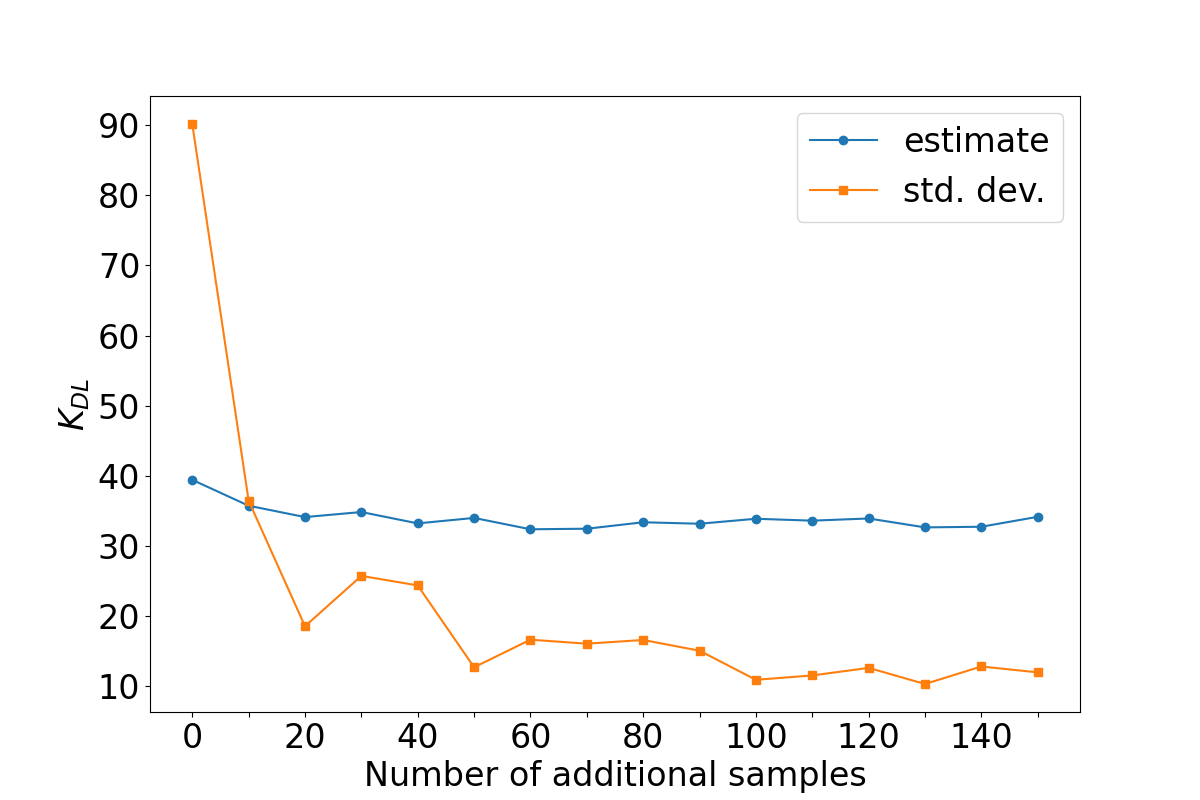}
        \caption{}
    \end{subfigure}
    \\
    \begin{subfigure}[b]{0.49\linewidth}
        \centering
        \includegraphics[width=\linewidth]{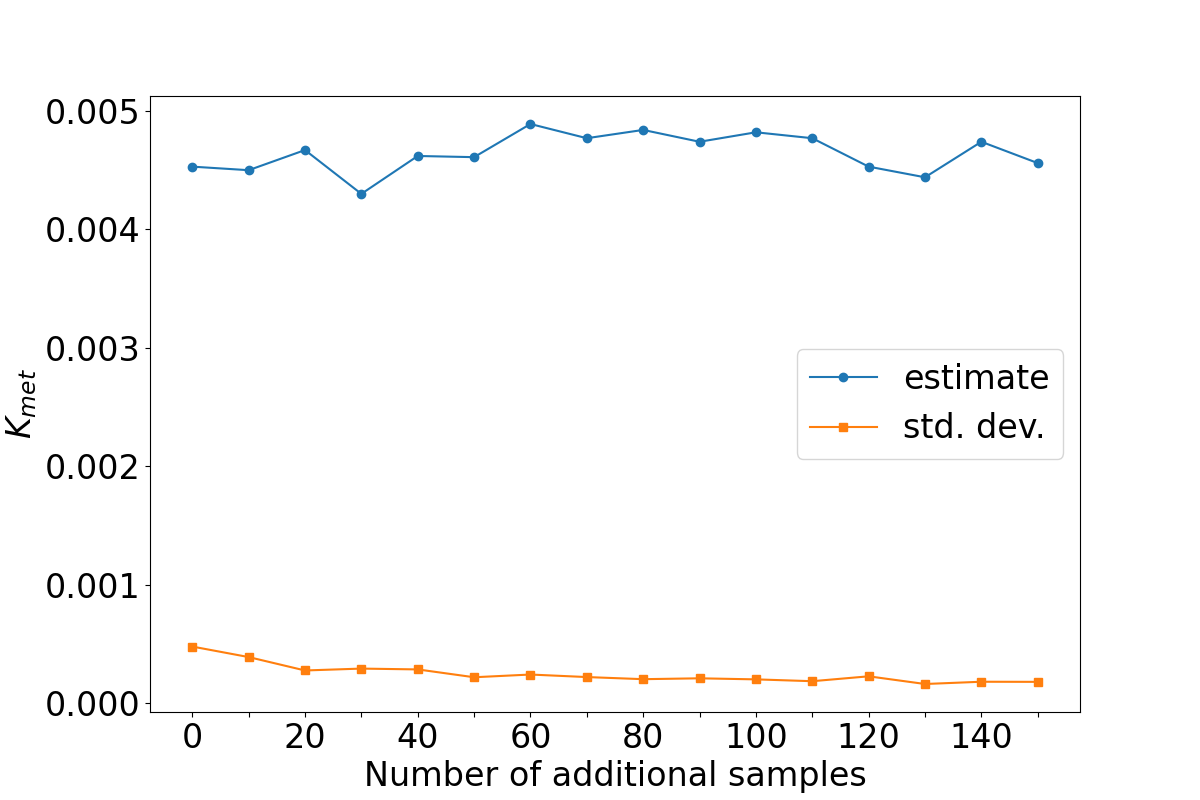}
        \caption{}
    \end{subfigure}
    \hfill
    \begin{subfigure}[b]{0.49\linewidth}
        \centering
        \includegraphics[width=\linewidth]{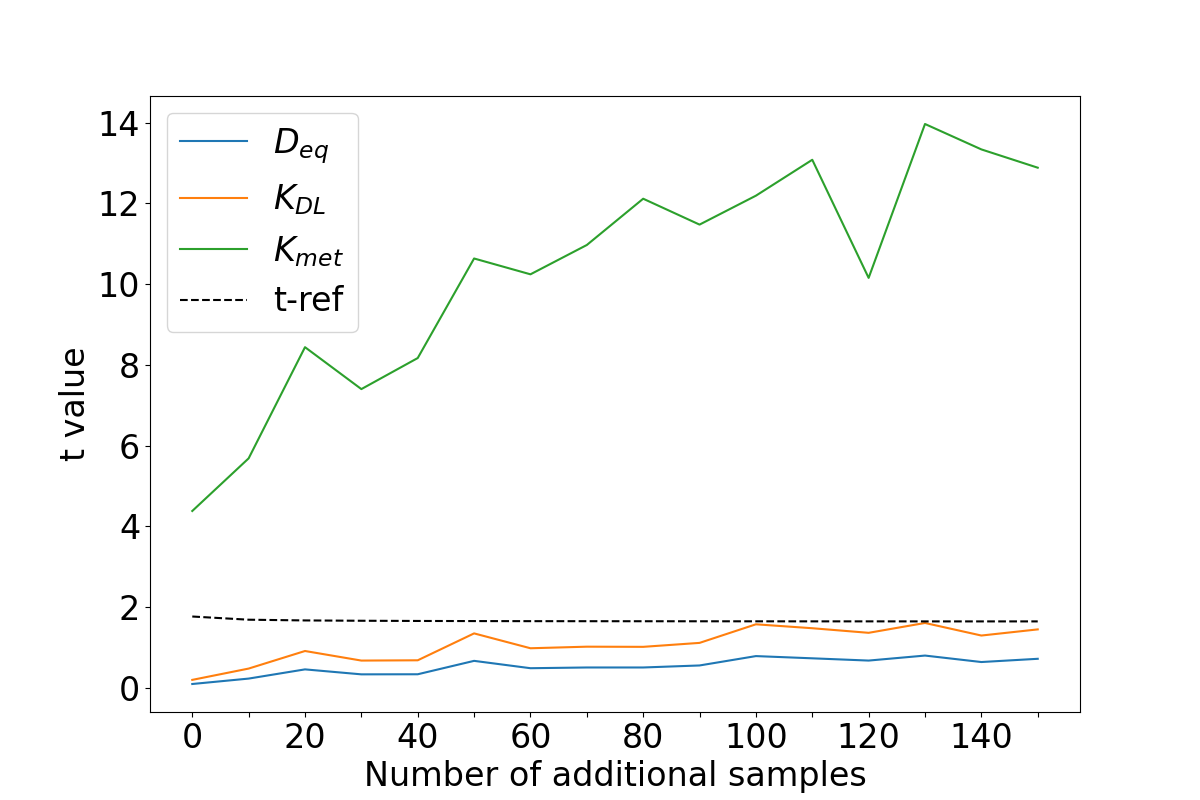}
        \caption{}
    \end{subfigure}    
    \caption{Data augmentation study for the diffusion-based model showing (a-c) the reduction in parametric uncertainty with the increased number of sampling points, and (d) the profile of t-test statistics. Dashed lines show the values obtained with the original dataset.}
    \label{fig:diff-no-budget-limit}
\end{figure}

Figure~\ref{fig:diff-no-budget-limit} shows the results of the analysis: the plots (a), (b), and (c) report the estimated value and standard deviation for $D_{eq}$, $K_{DL}$, and $K_{met}$ respectively, while in (d) the t-value profiles are plotted against the number of additional experimental points. 
It is interesting to observe that the uncertainty on both $D_{eq}$ and $K_{DL}$ drops significantly with the first 50 additional data point. After that point, the uncertainty keeps decreasing further, as shown by the increase in the t-values, but with a lower rate of improvement. The t-value of $K_{DL}$ approaches the threshold with more than 100 new points, while the t-value of $D_{eq}$ is well below the limit also under these conditions.

This study can provide directions to the experimenters around what they should mostly look at before consuming experimental resources, which is the added value of combining computational research with the experimental work in this application. Still, this is a hypothetical study under the assumption of almost unlimited budget, but then considerations must be take into account on the actual experimental budget and practical feasibility of collecting that many samples in a short time period, not to mention that the estimate are still not statistically good even in this theoretical scenario. 

This analysis demonstrates that, to solve the practical identifiability issues for $D_{eq}$ and $K_{DL}$, simply collecting more samples from the same type of bio-kinetic experiments is not enough. The modelers and experimentalists should then work together to evaluate other potential solutions, such as i) collecting data from different experiments that can allow an independent estimation of one of the two parameters, ii) correlate the problematic parameters to other physical and chemical properties of the crop-AI-formulation system, iii) reformulate the diffusion-based model.
In terms of future directions, the possibility of using less but very precise samples will also be explored, combining it with a parametric study on the variance model for measurement errors.

\section{Conclusions}
\label{sec:conclusions}
This paper presented the application of a systematic modeling framework to understand and characterize the process of foliar uptake of pesticides. Different models have been considered, namely compartmental and diffusion-based models, and compared within the proposed modeling approach.

The study aims to develop a model that can be used in practice, together with the experimental observations, to extract more information about the system defined by crop-active ingredient-product formulation and potentially predict the expected uptake for new systems. To achieve this goal it is crucial to include practical considerations during all the steps of the modeling procedure, first of all ensure that model parameters are identifiable.

The analyses conducted in this paper focus on the concept of a-posteriori identifiability, which has been tested via dynamic sensitivity profiles and by studying the correlation between the parameters. In silico data augmentation studies have been performed to evaluate the expected improvement in the estimates and to assess the possibility of solving identifiability issues by collecting additional samples.

The parameters of the compartmental model are practically identifiable, even though some of them (i.e. $k_{drop,loss}$ and $k_{drop,store}$) have a large uncertainty in the preliminary estimate. The data augmentation study has demonstrated that their statistical quality can be easily improved by adding a few data points collected at the most informative sampling times.

In the diffusion-based model, the parameter $K_{met}$ describing the metabolic rate of of AI consumption in the leaf is well identified, but practical identifiability issues arise between the diffusion coefficient in the leaf, $D_{eq}$, and the partition coefficient at the interface droplet-leaf, $K_{DL}$. These two parameters are totally anti-correlated and even with a large experimental budget the data would not be sufficient to precisely estimate them.
To address this issue, a co-operation between modelers and experimentalists is advised to evaluate potential solutions, such as i) collecting data from different experiments that can allow an independent estimation of one of the two parameters, ii) correlate the problematic parameters to other physical and chemical properties of the crop-AI-formulation system, iii) reformulate the diffusion-based model.

This study paves the way to further developments in the application of model-based design of experiments in a biological context characterized by high uncertainty in the experimental observations and in the model parameters. Further developments of the study will cover the uncertainty propagation from the model parameters to the predictions, coupled with the data augmentation study presented here.
Finally, the application of model-based design of experiments (MBDoE) in foliar uptake experiments is also part of future works, to validate the application of these techniques in a novel context.

\section*{Acknowledgments}
The authors gratefully acknowledge the support of the Department of Chemical Engineering, University College London, UK and Syngenta.

\section*{CRediT authorship contribution statement}
\textbf{Enrico Sangoi}: Writing – original draft, Writing – review and editing, Formal analysis, Investigation, Methodology, Conceptualization.
\textbf{Federica Cattani}: Writing – review and editing, Methodology, Supervision, Conceptualization.
\textbf{Faheem Padia}: Data curation. 
\textbf{Federico Galvanin}: Writing – review and editing, Methodology, Supervision, Conceptualization.

\bibliographystyle{elsarticle-harv} 
\bibliography{main}

\end{document}